\documentclass[aps,twocolumn,epsf,floats,pre,superscriptaddress,nofootinbib]{revtex4-1}

\usepackage{graphicx}
\usepackage{amsmath}
\usepackage{microtype}
\usepackage{tikz}
\usepackage[compat=1.1.0]{tikz-feynman}
\usepackage{commath}
\usepackage{enumitem}
\usepackage{amsfonts}
\usepackage{hyperref}
\usepackage{mathtools}
\usepackage{scrtime}
\usepackage[dont-mess-around]{fnpct}
\usepackage{import}
\usepackage{etoolbox}

\usetikzlibrary{external}
\immediate\write18{mkdir -p pgf-img}
\newcommand{\Ham}{\mathcal{H}}
\newcommand{\PP}{\mathbb{P}}
\newcommand{\II}{\mathbb{I}}
\newcommand{\tdelta}{\hat{\delta}}

\newcommand{\vv}{{\boldsymbol{v}}}
\newcommand{\hpsi}{\hat{\psi}}
\newcommand{\vpsi}{{\boldsymbol{\psi}}}
\newcommand{\hvpsi}{\hat{\vpsi}}
\newcommand{\hs}{\hat{s}}
\newcommand{\vs}{\boldsymbol{s}}
\newcommand{\hvs}{\hat{\vs}}
\newcommand{\ttheta}{\tilde{\theta}}
\newcommand{\vx}{\boldsymbol{x}}

\newcommand{\he}{\hat{e}}
\newcommand{\ve}{\boldsymbol{\he}}
\newcommand{\vk}{\boldsymbol{k}}
\newcommand{\tvk}{\tilde{\vk}}
\newcommand{\vq}{\boldsymbol{q}}
\newcommand{\tvq}{\tilde{\vq}}
\newcommand{\vh}{\boldsymbol{h}}
\newcommand{\tvh}{\tilde{\vh}}
\newcommand{\vp}{\boldsymbol{p}}
\newcommand{\tvp}{\tilde{\vp}}
\newcommand{\vnabla}{{\boldsymbol{\nabla}}}
\newcommand{\iu}{\text{i}}
\newcommand{\eu}{\text{e}}

\newcommand{\pder}[2]{\frac{\partial #1}{\partial #2}}
\newcommand{\dder}[2]{\frac{\delta #1}{\delta #2}}
\newcommand{\vecc}[1]{\boldsymbol{#1}}

\begin{document}

\title{Dynamical renormalization group for mode-coupling field theories\\ with solenoidal constraint}

\author{Andrea Cavagna}
\affiliation{Istituto Sistemi Complessi, Consiglio Nazionale delle Ricerche, UOS Sapienza, 00185 Rome, Italy}
\affiliation{Dipartimento di Fisica, Universit\`a\ Sapienza, 00185 Rome, Italy}
 
\author{Luca Di Carlo}
\affiliation{Dipartimento di Fisica, Universit\`a\ Sapienza, 00185 Rome, Italy}
\affiliation{Istituto Sistemi Complessi, Consiglio Nazionale delle Ricerche, UOS Sapienza, 00185 Rome, Italy}

\author{Irene Giardina}
\affiliation{Dipartimento di Fisica, Universit\`a\ Sapienza, 00185 Rome, Italy}
\affiliation{Istituto Sistemi Complessi, Consiglio Nazionale delle Ricerche, UOS Sapienza, 00185 Rome, Italy}
\affiliation{INFN, Unit\`a di Roma 1, 00185 Rome, Italy}

\author{Tomas S. Grigera}
\affiliation{Instituto de F\'\i{}sica de L\'\i{}quidos y Sistemas Biol\'ogicos CONICET -  Universidad Nacional de La Plata,  La Plata, Argentina}
\affiliation{CCT CONICET La Plata, Consejo Nacional de Investigaciones Cient\'\i{}ficas y T\'ecnicas, Argentina}
\affiliation{Departamento de F\'\i{}sica, Facultad de Ciencias Exactas, Universidad Nacional de La Plata, Argentina}

\author{Giulia Pisegna}
\affiliation{Dipartimento di Fisica, Universit\`a\ Sapienza, 00185 Rome, Italy}
\affiliation{Istituto Sistemi Complessi, Consiglio Nazionale delle Ricerche, UOS Sapienza, 00185 Rome, Italy}

\author{Mattia Scandolo}
\affiliation{Dipartimento di Fisica, Universit\`a\ Sapienza, 00185 Rome, Italy}
\affiliation{Istituto Sistemi Complessi, Consiglio Nazionale delle Ricerche, UOS Sapienza, 00185 Rome, Italy}

\begin{abstract}
The recent inflow of empirical data about the collective behaviour of strongly correlated biological systems has brought field theory and the renormalization group into the biophysical arena. Experiments on bird flocks and insect swarms show that social forces act on the particles' velocity through the generator of its rotations, namely the spin, indicating that mode-coupling field theories are necessary to reproduce the correct dynamical behaviour. Unfortunately, a theory for three coupled fields - density, velocity and spin - has a prohibitive degree of intricacy. A simplifying path consists in getting rid of density fluctuations by studying incompressible systems. This requires imposing a solenoidal constraint on the primary field, an unsolved problem even for equilibrium mode-coupling theories. Here, we perform an equilibrium dynamic renormalization group analysis of a mode-coupling field theory subject to a solenoidal constraint; using the classification of Halperin and Hohenberg, we can dub this case as a solenoidal Model G. We demonstrate that the constraint produces a new vertex that mixes static and dynamical coupling constants, and that this vertex is essential to grant the closure of the renormalization group structure and the consistency of dynamics with statics. Interestingly, although the solenoidal constraint leads to a modification of the static universality class, we find that it does not change the dynamical universality class, a result that seems to represent an exception to the general rule that dynamical universality classes are narrower than static ones. Our results constitute a solid stepping stone in the admittedly large chasm towards developing an off-equilibrium mode-coupling theory of biological groups.
\end{abstract}

\maketitle

\section{Introduction}

The phenomenon whereby systems with very different microscopic details have similar critical behaviour, known as universality, finds an elegant explanation in the context of the Renormalization Group (RG) \cite{wilson1971renormalization1,wilson1971renormalization2}.
As a system approaches a critical point, i.e. a second order phase transition, it exhibits emergent collective phenomena on increasingly wider length-scales, which leads to the presence of a diverging correlation length.
In this regime, a set of phenomenological scaling laws for near-critical systems have been proposed \cite{widom1965equation,kadanoff1966scaling} in order to describe the critical behaviour through a set of \textit{critical exponents} ruling the divergence of thermodynamic quantities with respect to the correlation length.
In the RG framework, these scaling laws naturally arise as a consequence of the diverging correlation length, thanks to the flow in the parameter space generated by iterating an RG transformation many times.
Through the concept of attractive fixed points of the RG flow, it is possible to prove that systems sharing only few general properties, as symmetries, dimensionality and range of the interactions, develop the same large-scale behaviour near the critical point, hence belonging to the same universality class.
Moreover, the RG gives a constructive method to compute the critical exponents \cite{goldenfeld_lectures_1992}, making it one of the most powerful tools of the theory of critical phenomena, broadly employed to characterise both static \cite{pelissetto2002critical} and dynamic \cite{HH1977} universality classes in equilibrium systems.

When we shift our attention from equilibrium to non-equilibrium phenomena, the study of emergent collective dynamics is of fundamental interest also in biological systems, in which the response to external perturbations is more efficient when individuals behave collectively \cite{cresswell1994flocking, roth2006determinants}.
The existence of collective behaviour in biological systems is strictly related to the presence of  scale-free correlations among the system's components, as experimentally observed in bird flocks \cite{cavagna2010scalefree}, sheep herds \cite{ginelli2015herds} and bacterial clusters \cite{zhang2010collective}, in strong similarity with emergent phenomena studied in condensed matter, as superfluidity, superconductivity and ferromagnetism.
The experimental discovery of scale-free correlations provides some justification to the theoretical investigation of collective behaviour in biological systems using the tools developed for critical phenomena, including of course the renormalization group.

The early efforts in the study of collective phenomena in active biological systems focused on the ordered phase, namely in those groups exhibiting net group motion, the most prominent example being represented by bird flocks \cite{tonertu1995,tonertu1998}. More recently it has become experimentally evident that also systems in their disordered phase, as natural swarms of insects, display collective phenomena even in absence of collective order \cite{attanasi2014collective}. Moreover, within natural swarms both static and dynamic scaling laws have been found to hold \cite{attanasi2014collective,cavagna2017swarm}.
Since scaling laws are fundamental features of near-critical systems, a renormalization group approach to swarming behaviour seems particularly appealing.

Dynamic critical systems not only have a diverging correlation length, but also a diverging relaxation time \cite{cardy1996scaling}.
According to the dynamic scaling hypothesis, the correlation length $\xi$ and the collective (i.e. zero wavelength) relaxation time $\tau$ are linked by the relation $\tau\sim\xi^z$, where $z$ is the dynamic critical exponent ruling the dynamical relaxation of the system. Experiments show that the dynamic behaviour of natural swarms is characterized by a critical exponent $z\sim1$ \cite{cavagna2017swarm}, which is very far from the value $z\sim2$ typical of the simplest dissipative ferromagnetic models, generically represented by Model A in the famous classification of Halperin and Hohenberg \cite{HH1977}; this is somewhat problematic, as the earliest theories for active biological systems are essentially a mix of Model A ferromagnets and Navier-Stokes equations \cite{tonertu1995}. Finding what are the correct dynamical equations that provide a value of the dynamical critical exponent $z$ equal to the experimental value represents a very important step, not only towards understanding collective behaviour in natural swarms (which will be our focus here), but more broadly in the effort to prove that statistical field theory is a predictive tool in biophysics.

There are two possible mechanisms that can explain a drop of the value of the dynamical critical exponents in natural swarms compared to that of standard ferromagnetic systems in the Model A class. The first possibility is activity, namely the fact that individuals are self-propelled, which leads to a time-dependent network of interactions between the particles' velocities, thus driving the system out of equilibrium. Activity gives rise to density fluctuations, thus meaning that in addition to the velocity field $\vecc{v}\left(\vx,t\right)$, also the density field $\rho\left(\vx,t\right)$ must be included in the hydrodynamic description of the system \cite{tonertu1995,tonertu1998}. Broadly speaking, one can say that the feedback between velocity and density fluctuations is responsible for a more efficient transport of information within the system, which may reflect into a lower value of $z$. Indeed, theoretical RG calculations in presence of activity have been performed in the incompressible case, showing that $z$ is indeed lower than its equilibrium value, leading the exponent from $z\sim2$ to $z\sim1.7$ in $d=3$ dimensions \cite{chen2015critical}; and yet this is still far from the experimental value $z\sim1$.

The second factor that may lower the value of $z$, and that has been shown to be relevant in the description of both insect swarms and bird flocks, is the presence of inertial dynamics, namely a coupling between the velocity field  $\vecc{v}\left(\vx,t\right)$ and the generator of its rotations, known as the spin $\vecc{s}\left(\vx,t\right)$ \cite{attanasi2014information,cavagna2017swarm}. In equilibrium systems, as for example planar ferromagnets, superfluids  and quantum antiferromagnets (respectively Model E, F and G in the classification of \cite{HH1977}) this coupling (and the conservation law generated by it) radically alters the mechanism of information propagation throughout the system and it also lowers the dynamical critical exponent, which becomes $z=1.5$ in $d=3$ \cite{cavagna2019short,cavagna2019long}. Even in this case, though, we are quite far from the experimental value, but then it seems likely that a calculation that keeps into account both the off-equilibrium active nature of swarms {\it and} the mode-coupling nature of their interaction, may provide a value of the dynamical critical exponent close to the experimental one, $z\sim 1$. The bad news is that a full-fledged out of equilibrium field theory taking into account both these features must contain a density field $\rho\left(\vx,t\right)$, a velocity field $\vecc{v}\left(\vx,t\right)$ and a spin field $\vecc{s}\left(\vx,t\right)$, with a large number of couplings between them and a near-prohibitive level of intricacy. Therefore, to make theoretical progress, one needs some simplifying assumptions.

The first and most natural simplification is that of incompressibility. Incompressibility in the RG study of the Navier-Stokes equations was first used in the seminal paper by Forster, Nelson and Stephen \cite{FNS1977}, while it was used first in the context of dynamical field theories for active matter in \cite{chen2015critical}. This hypothesis seems quite reasonable both in natural insect swarms and bird flocks, because experiments show that density fluctuations are indeed very limited in these systems, and they do not seem to be a crucial ingredient in their collective behaviour \cite{attanasi2014collective,mora2016local}. Incompressibility is obtained by enforcing a solenoidal constraint on the velocity field, that is by adding to the dynamics the condition, $\vnabla\cdot\vv=0$. This allows to neglect the (constant) density field, whose fluctuations are suppressed thanks to the solenoidal constraint. Incompressible swarms in the fully dissipative case (i.e. without velocity-spin coupling) have been studied in \cite{chen2015critical}, but we need to move to the mode-coupling case, of velocity and spin, and here some new problems arise. In the case of a theory with one single field (the velocity), the solenoidal constraint is simply enforced by projecting the force acting on the velocity  onto the direction orthogonal to the wave vector $\boldsymbol k$ in Fourier space \cite{FNS1977,chen2015critical}; in this way, the constraint is conserved along the time trajectories. But in systems with mode-coupling dynamics incompressibility cannot be achieved in this simple way, as the forces now act on the spin, rather than on the velocity, and the spin is not subject to the solenoidal constraint! Hence we are in a paradoxical situation: the assumption that was supposed to radically simplify the calculation, reducing from three to two the number of fields, creates itself a new challenges.

Instead of giving up the crucial simplification of incompressibility, we choose to try and solve the problem of how to impose a solenoidal constraint in a mode-coupling theory; and because this new problem has nothing to do with the complications due to off-equilibrium dynamics, we tackle the problem in the case of a zero-activity equilibrium system, namely a system living on a fixed-network. Indeed, surprising as this may seem, the problem of how to impose the solenoidal constraint on a mode-coupling theory has never been investigated so far. Once we will have solved this problem, we will be in a far better shape to push the calculation out of equilibrium, still retaining the incompressibility condition. Notice that in absence of activity it does not make much sense to insist in calling the primary field ``velocity". Hence, we will use as our primary field $\vpsi$, the field describing the average direction of motion in the vanishing speed limit. The virtue of this approach is that any future calculation performed in the active case must have the present calculation as a limit in the zero-speed case.

To summarize, in the present work we will derive and study the coarse-grained dynamic equations for a critical field $\vpsi$ which is coupled to the generator of its rotations, $\vs$, in the presence of the solenoidal constrain, $\vnabla\cdot\vpsi=0$. We shall refer to this theory as Solenoidal Model G (SMG), because in absence of the solenoidal constraint this field theory is known in literature as Model G \cite{HH1977}: this classic model for equilibrium quantum antiferromagnets will therefore be our stepping stone towards a theory for off-equilibrium natural swarms. We will show that the solenoidal constraint leads to the emergence of an additional non-linear term in the equation of motion of the spin, as a consequence of the suppression of the $\vpsi$ mode longitudinal to the wave-vector $\vk$. We will also prove that the equations we obtain are eigenstates of the renormalization group, by calculating perturbative corrections up to one loop. Moreover, the dynamic behaviour will be shown to be consistent with the static behaviour of a ferromagnet with dipolar interactions \cite{aharony1973critical,fisher1973dipolar}. Finally, we will find as dynamic critical exponent $z=\frac{d}{2}$, leading to the same dynamic universality class as the unconstrained theory \cite{cavagna2019short,HH1977}. Therefore, while incompressibility affects the static universality class, it does not modify the dynamic universality class, suggesting that the critical behaviour of homogeneous systems is not affected by imposing incompressibility.

In \autoref{sec:bio} we will review the biological origin of the mode-coupling dynamics and the reason for which incompressible constraint is imposed. In \autoref{sec:theory} the coarse-grained theory will be derived, starting form the static Hamiltonian and the Poisson-bracket relations between the various quantities. In \autoref{sec:RG} we will derive one-loop corrections to the couplings of the theory and verify that the equations of motion are self-consistent. Self-consistency is proven by showing that no new RG-relevant interaction is generated and that the dynamic renormalization reproduces the correct static behaviour. Finally, in \autoref{nostromo} we will derive the dynamical critical exponent.


\section{Biophysical background}
\label{sec:bio}

\subsection{Dynamic scaling}

The spatio-temporal statistical behavior of a collective system can be described through the connected space-time correlation function of the velocity, $C(\boldsymbol r,t)$, which expresses how much the velocity fluctuations of some individual in the system influence and are influenced by the velocity fluctuations of another individual at distance $\boldsymbol r$ and after a time $t$ \cite{cavagna2018correlations}. For $t=0$ this is just the static (i.e. equal times) correlation function, whose decay range defines the {\it correlation length}, $\xi$. Conversely, for $r=0$ and $t\neq 0$, this is the singe-particle auto-correlation function. In general, when both $r$ and $t$ are non-zero, the correlation function is quite a loaded concept, entailing the fundamental relationship between relaxation in time and relaxation in space. In the case of critical systems, such relationship acquires a particularly illuminating form, which goes under the name of {\it dynamic scaling} \cite{Ferrell1967,HH1967scaling}. Dynamic scaling states that the dynamic correlation function $C$, when expressed as a function of wave-vector and frequency, takes the following simplified scaling form, 
\begin{equation}
C\left(\vk,\omega;\xi\right)=C_0\left(\vk;\xi\right)\,F\left(\frac{\omega}{\omega_k},k\xi\right)\\
\end{equation}
where $\xi$ is the correlation length and where the {\it static} correlation function $C_0$ has in turns the scaling form,
\begin{equation}
C_0(k,\xi)=k^{2-\eta}F_0\left(k\xi\right)
\end{equation}
and where the characteristic frequency at scale $k$ is given by,
\begin{equation}
\omega_k=k^z\Omega\left(k\xi\right)
\end{equation}
In the relations above, $\Omega$, $F_0$ and $F$ are well-behaved scaling functions, whose explicit form is inessential to capture the gists of dynamic scaling \cite{Ferrell1967,HH1967scaling}; $\eta$ is the critical exponent for the static correlation function (normally called anomalous dimension \cite{binney_book}). The fundamental meaning of dynamic scaling is that in critical systems space and time do not scale independently from each other, but they are linked by the dynamic critical exponent $z$. The space-time correlation function has a very simple form, as its whole spatio-temporal dependence goes through the product $k\xi$. When $\vk=0$, the collective relaxation time $\tau$ of the system is linked to the correlation length $\xi$ through the relation,
\begin{equation} 
\tau \simeq\omega_{\vk=0}^{-1}\sim\xi^z
\end{equation}
This phenomenon is known as \textit{critical slowing down} and it represents a consequence of the fact that the time $\tau$ needed to decorrelate a spatially correlated region grows with the region's size, namely with the correlation length $\xi$, making the latter the only relevant scale also at a dynamic level.

Notably, natural swarms of insects have been found to obey dynamic scaling \cite{cavagna2017swarm}, with a dynamical critical exponent $z$ quite close to $1$ ($z=1.2$, obtained by looking at the collapse of the correlation functions \cite{cavagna2017swarm}, is the safest determination). On one hand, because the validity of scaling laws is one of the hallmarks of critical systems, the experimental evidence of static and dynamic scaling  in natural swarms \cite{attanasi2014collective,cavagna2017swarm} suggests that the swarming phase can be theoretically described as a near-ordering phase of classic ferromagnetic theories. On the other hand, the value $z\sim1$ is definitely anomalous, as the standard exponents in Model A class of statistical systems \cite{HH1977} is $z\sim 2$, with small corrections at two loops. This state of affairs suggest that a critical dynamical universality class, different from Model A, must be found for natural swarms. This new universality class will need to take into account the two main features that affect the collective behaviour of these systems, namely activity and mode-coupling dynamics. Let us illustrate the first factor through the most venerated hydrodynamic theory describing active matter.

\subsection{The hydrodynamic theory of Toner and Tu}

In most collective biological systems, from bacterial clusters up to insect swarms, bird flocks and animals herds, individuals are self-propelled thanks to their ability to convert energy into systematic movement \cite{schweitzer2003brownian}. The presence of metabolic processes, through which an energy supply from the environment is provided, guarantees a constant speed. Activity means that the interaction network over which the velocities of the individuals are interacting with each other evolves in time, representing the principal non-equilibrium feature of these systems.

The hydrodynamic theory developed by Toner and Tu \cite{tonertu1995,tonertu1998}, based on the discrete model proposed by Vicsek and collaborators \cite{vicsek1995novel}, paved the way to a theoretical analysis of collective behaviours in the presence of activity.
Together with the velocity field $\vv$, also the density number field $\rho$ is an hydrodynamic variable of the system, since activity allows local density fluctuations.
The velocity field obeys to a dynamical behaviour that is a crossover between that of a  Landau-Ginzburg $O(n)$ model, reflecting the presence of an effective alignment among the animals velocities, and a Navier-Stokes dynamics, reflecting the fact that velocity and density are coupled as in a standard fluid; the density evolves according to a continuity equation, since the total number of individuals is assumed to be fixed.
The essential terms of the Toner and Tu (TT) theory are given by \cite{ramaswamy2010mechanics},
\begin{gather}
\pder{\vv}{t}+ \gamma_v \left(\vv\cdot\vnabla\right)\vv+\dots=-\Gamma\dder{\Ham}{\vv}+\vnabla P+\vecc{\theta}\label{eq:TTv}\\
\pder{\rho}{t} =-\vnabla\cdot \left(\rho\vv\right)\label{eq:TTp}
\end{gather}
In Eq.~\eqref{eq:TTv}, the r.h.s. encloses Model A dynamics \cite{HH1977}, with the gaussian random white noise $\vecc{\theta}$ and the hamiltonian force $-\dder{\Ham}{\vv}$, to which a pressure force is added contrasting an infinite compressibility.
Here the effective hamiltonian $\Ham$ takes the Landau-Ginzburg structure, namely
\begin{equation}
\Ham=\int d^d x \left[\frac{1}{2}(\vnabla\vv)^2+\frac{r_0}{2}\vv\cdot\vv+\frac{u_0}{4}\left(\vv\cdot\vv\right)^2\right]
\end{equation}
Depending on the value of $r_0$, two phases can be identified: an ordered phase for $r_0 < r_c$ and a disordered phase for $r_0>r_c$, where $r_c$ is the critical value of the parameter $r$.
The main difference with equilibrium dynamics is represented by the $\gamma_v$ term on the l.h.s. of Eq.~\eqref{eq:TTv}, which is the advection term%
\footnote{Since Galilean invariance is violated, we cannot require $\gamma_v=1$. Moreover, the ellipsis in Eq.~\eqref{eq:TTv} denote other two advection-like terms, involving one $\vnabla$ and two $\vv$, allowed by symmetries}
typical of the Navier-Stokes structure.
 
Since its development, the TT theory has been able to explain many fundamental aspects of collective behaviours in active systems, from the existence of an ordered phase also in $2$ dimensions, to the presence of linear sound modes \cite{tonertu1995,tu1998sound}.
The success of this remarkable theoretical effort is due to the minimal description given by the TT equations of motion, which tie together activity and effective alignment.
However, experimental evidence has shown how this theory is not able to fully describe some aspects of biological systems', in particular information propagation in flocks \cite{attanasi2014information}, and dynamical correlation functions in swarms \cite{cavagna2017swarm}.
These discrepancies arise as consequence of the structure of Eq.~\eqref{eq:TTv}, which is essentially that of an overdamped Langevin equation for the velocity, 
\begin{equation}
D_t \vv = - \partial_\vv \Ham
\end{equation}
Moreover, RG analysis of the near-ordering dynamics of the TT theory in the incompressible case predictss a dynamical critical exponent $z\simeq1.7$ for $d=3$ \cite{chen2015critical}, significantly different from the value $z\sim1$ observed in natural swarms.

\subsection{Restoring inertia: the mode-coupling theory}

The first hints about the necessity of restoring an underdamped dynamics came from experiments on bird flocks. In order to explain the information propagation dispersion law within turning flocks, the generator of the rotations of the velocity vector has to be conserved during the turn \cite{attanasi2014information}. In analogy with quantum mechanics, this generator of the rotations in the \textit{internal} space of the velocity, which must not be confused with the generator of spatial rotations (the standard angular momentum), takes the name of \textit{spin}.

A second piece of empirical evidence that inertial dynamics was missing from the original Vicsek model and Toner-Tu theory came from the calculation of the dynamical correlation functions in real swarms \cite{cavagna2017swarm}. The overdamped dynamics for the velocity typical of the Vicsek model and TT theory generates a classic exponential decay of the dynamical correlation functions; however, in natural swarms this is not the case, and the correlation function displays a non-exponential form typical of underdamped inertial systems. This feature too can be included at the theoretical level by incorporating in the dynamics the generator of the rotations of the velocity.

The discrete model describing this kind of spin-velocity mode-coupling behaviour is known as the Inertial Spin Model (ISM) \cite{cavagna2015flocking}, and it is characterized by an underdamped structure for the equations of motion; where the alignment force exerted on the focal particle by its neighbours does not act directly on the velocity, but it is mediated by the generator of rotations, thus restoring an inertial behaviour.

Adding a conservation law leads to a modification of the hydrodynamic behaviour of a system, which now must include the dynamics of the spin density and the effects of a conserved total spin on the other variables.
In the present case, the conservation of the spin leads a mode-coupling between the order parameter and the spin density field.
This structure is not new to physical systems: this is what happens in the case of planar magnets, superfluid helium and isotropic anti-ferromagnets (Models E, F and G in \cite{HH1977} respectively).
The equations of motion for these models, in terms of the order parameter $\vpsi$ and the spin $\vs$, take the following form for a $3$ dimensional order parameter (Model G)
\begin{align}
\pder{\vpsi}{t}&=-\Gamma\dder{\Ham}{\vpsi}+g\, \vpsi \times \dder{\Ham}{\vs} + \vecc{\theta}\label{eq:psiMG}\\
\pder{\vs}{t}&=\lambda \nabla^2\dder{\Ham}{\vs}+g\,\vpsi \times \dder{\Ham}{\vpsi}+\vecc{\zeta}\label{eq:sMG}
\end{align}
where $\vecc{\theta}$ and $\vecc{\zeta}$ are gaussian random white noises, while the Hamiltonian has the usual Landau-Ginzburg form for $\vpsi$, plus a Gaussian non-interacting kinetic term in the spin,
\begin{equation}
\Ham=\int d^d x \left[\frac{1}{2}\left(\vnabla\vpsi\right)^2+\frac{r_0}{2}\vpsi\cdot\vpsi+\frac{u_0}{4}\left(\vpsi\cdot\vpsi\right)^2+\frac{\vs\cdot\vs}{2}\right]
\label{eq:Ham}
\end{equation}
In Eq.~\eqref{eq:psiMG} and \eqref{eq:sMG} the two terms proportional to $g$, namely $\partial_t \psi \sim \delta_s \Ham$ and $\partial_t s \sim \delta_\psi \Ham$, represent the Hamiltonian conservative dynamics enforced by the Poisson-braket relation
\begin{equation}
\left\{s_\alpha,\psi_\beta\right\}=g_0\, \epsilon_{\alpha\beta\gamma}\psi_\gamma
\end{equation}
where $\epsilon_{\alpha\beta\gamma}$ is the Levi-Civita antisymmetric symbol, meaning that $\vs$ is rotating $\vpsi$.
Moreover, the equation of motion of $\vs$ can be written in the form of a continuity equation, reflecting the fact that due to the rotational symmetry, the total spin $\vecc{S}(t)=\int d^d x \ \vs(x,t) $ is conserved%
\footnote{In the biophysical context, the spin is not strictly conserved. Thus, the presence of a dissipative term $\partial_t \vs\sim-\eta\delta_s\Ham$ is allowed in Eq.~\eqref{eq:sMG}. In this system a crossover between a conservative and a dissipative dynamics has been observed \cite{cavagna2019short,cavagna2019long}, in which the former regulates the behaviour of finite-size systems as swarms. As we will see later on, the solenoidal constraint will violate the spin conservation, but does not generate a dissipative behaviour, thus allowing us to work at $\eta=0$}.

\subsection{Incompressible flow and fixed network assumptions}

In order to join the two ingredients described in the last two sections, namely activity and inertia (i.e. mode-coupling dynamics) into one single
dynamical field theory, one needs to write equations for three coupled fields - density, velocity and spin - with several non-linear couplings among them, giving rise to a diagrammatic RG proliferation that it is impossible to keep under control without any simplifying assumption. One such simplification is that of incompressibility. The huge advantage of working under the hypothesis of incompressibility is that the density field drops out the theoretical description \cite{FNS1977,chen2015critical}, thus reducing the number of fields that have to be studied from three to two, hence dramatically simplifying the theoretical investigation. One may worry that incompressibility introduces some non-local interactions in the system, then potentially changing the critical exponents with respect to the compressible case. However, numerical simulations of the standard {\it compressible} theory have found the dynamic scaling exponent to be in perfect agreement with theoretical RG results for {\it incompressible} theory \cite{cavagna2020equilibrium}, thus indicating that the effects of activity on the dynamical critical exponent are independent from whether incompressibility is enforced or not. This is very useful indeed, as it means that incompressibility can be used as a simplifying tool, without changing critical dynamics.

A second remark about incompressibility is in order. In the standard compressible theory, the presence of density fluctuations in the system, as a consequence of activity, makes the phase transition of Vicsek-like models a first-order (i.e. discontinuous) transition \cite{gregoire2004}. In incompressible systems, however, the absence of density fluctuations makes the phase transitions always second-order (continuous) \cite{chen2015critical}, thus scaling behaviour is observed at all sizes and all the complications about phase separation are avoided. Of course, this is only reasonable because in real natural swarms of insects (as in bird flocks), density fluctuations are negligible and play no determinant role in ruling the collective dynamics of the system \cite{attanasi2014collective,attanasi2014finite}. Hence, incompressibility is not only a simplifying hypothesis, but also a sound theoretical description of actual empirical data.

Thanks to the continuity equation \eqref{eq:TTp}, the condition of incompressibility  reduces to imposing a solenoidal constraint on the velocity field,
\begin{equation}
\vnabla\cdot\vv=0
\label{eq:inc-v-realspace}
\end{equation}
As we shall see, the introduction of a solenoidal constraint on the primary field within a mode-coupling dynamics is far from trivial. The only encouraging thing is that the complications arising are completely unrelated to activity, so that it seems sound to first find a conceptually consistent way to impose the solenoidal constraint on a theory at equilibrium, and then to use that results to make progress off-equilibrium in the future. Therefore, in this work activity will be neglected by assuming the adjacency network to be fixed in time. Since in the microscopic description of Vicsek-like systems each individual has a fixed speed $v_0$, the fixed-network approximation can be formally seen as a limit where $v_0$ vanishes, which is equivalent to \textit{freeze} the position of each particle.
While in this limit the velocity field acquires a singular behaviour, since no particle is actually moving, the coarse-grained direction of motion $\vpsi$, defined by the relation
\begin{equation}
\vv\left(\vx,t\right)=v_0\, \vpsi\left(\vx,t\right)
\label{eq:VvsPSI}
\end{equation}
still has a smooth behaviour when $v_0\to0$, and therefore represents the ideal candidate to be the order parameter in the fixed network approximation.
Since the solenoidal constraint given by Eq.~\eqref{eq:inc-v-realspace} holds also for $\vpsi$ at any finite value of $v_0$, it must hold also in the fixed network approximation. 


\section{The solenoidal mode-coupling theory}\label{sec:theory}

The first problem we have is that the presence of the solenoidal constraint,
\begin{equation}
\vnabla\cdot\vpsi=0
\end{equation}
forces the order parameter to have the same dimensionality as the space in which the theory is defined.
Thus, since in the following sections we will need to perform an RG expansion near $d=4$, we must work with the generalization of Model G to arbitrary dimensions, known as the Sasvari-Schwabl-Szepfalusy (SSS) model \cite{SSS1975,SSS1977}.\footnote{
Given that we will be calculating the properties of the Solenoidal Sasvari-Schwabl-Szepfalusy model, strictly speaking we should use the nomenclature ``SSSS model"; however, chiefly for aesthetic reasons, we prefer not to do that, and we will rather stick to Solenoidal Model G (SMG) even in generic dimension $d$.}
In this model, the spin $\vs$ is an anti-symmetric $d\times d$ matrix, with one independent component for each possible plane around which a rotation can be performed.
The Poisson-bracket relation between $\vs$ and $\psi$ becomes,
\begin{equation}
\left\{s_{\alpha\beta}\left(\vx\right),\psi_\gamma\left(\vx'\right)\right\}=2 g_0\, \II_{\alpha\beta\gamma\rho} \psi_\rho \left(\vx\right)\delta^{\left(d\right)}\left(\vx-\vx'\right)
\label{eq:poisson}
\end{equation}
where repeated greek-letter indices are intended to be summed, if not otherwise indicated, and the tensor $\II$ is the identity tensor in the space of $d\times d$ anti-symmetric matrices,\footnote{
The factor $\frac{1}{2}$ in the definition of $\II$ arises as a consequence of the fact that, when $\vs$ is represented as an anti-symmetric matrix, each independent component appears twice.} given by
\begin{equation}
\II_{\alpha\beta\gamma\rho}=\frac{1}{2} \left(\delta_{\alpha\gamma}\delta_{\beta\rho}-\delta_{\alpha\rho}\delta_{\beta\gamma}\right)
\end{equation}
The static properties of the SSS model are fully determined by its effective Hamiltonian, which reflects the fact that $\vpsi$ is a critical field with $O\left(d\right)$ symmetry group and local interactions, having the static critical properties of the Landau-Ginzburg universality class, while $\vs$ is a non-critical massive field.
The effective Hamiltonian of the system takes the following natural generalization
\begin{widetext}
\begin{equation}
\Ham=\int d^d x \left[\frac{1}{2} \left(\partial_\alpha\psi_\beta\right) \left(\partial_\alpha\psi_\beta\right) + \frac{r_0}{2} \psi_\alpha\psi_\alpha + \frac{u_0}{4} \left(\psi_\alpha\psi_\alpha\right)^2 + \frac{s_{\alpha\beta}s_{\alpha\beta}}{4}\right]
\label{eq:HamSSS}
\end{equation}
The equations of motion of the critical SSS model can be derived starting from the Poisson-bracket relation \eqref{eq:poisson} and the effective hamiltonian \eqref{eq:HamSSS} by following a procedure coming from the works of Mori \textit{et al.} \cite{mori1974new} and Zwanzig \cite{zwanzig1961memory} (see \cite{ma1975critical} and \cite{frey1994critical}).
These equations are given by
\begin{align}
\pder{\psi_\alpha}{t}&=-\Gamma_0 \dder{\Ham}{\psi_\alpha}+g_0 \dder{\Ham}{s_{\alpha\beta}}\psi_\beta +\theta_\alpha\label{eq:psiSSS}\\
\pder{s_{\alpha\beta}}{t}&=\lambda_0\nabla^2 \dder{\Ham}{s_{\alpha\beta}} + 2 g_0 \II_{\alpha\beta\gamma\nu}\,\psi_\gamma \dder{\Ham}{\psi_\nu}+\zeta_{\alpha\beta}\label{eq:sSSS}
\end{align}
where $\theta$ and $\zeta$ are two gaussian white noises, with variance
\begin{align}
\langle\theta_\alpha\left(\vx,t\right)\theta_\beta\left(\vx',t'\right)\rangle & = 2 \Gamma_0 \delta_{\alpha\beta} \delta^{(d)}\left(\vx-\vx'\right)\delta\left(t-t'\right)\\
\langle\zeta_{\alpha\beta}\left(\vx,t\right)\zeta_{\gamma\nu}\left(\vx',t'\right)\rangle & = - 4 \lambda_0 \II_{\alpha\beta\gamma\nu} \nabla^2 \delta^{(d)}\left(\vx-\vx'\right)\delta\left(t-t'\right)
\end{align}
\end{widetext}
The $\lambda_0\nabla^2$ terms in the stochastic parts of the dynamics of $\vs$ ensure that the spin is conserved.

\subsection{Static critical behaviour of the solenoidal theory}\label{sec:static}

We will now review, for the benefit of the reader, the classic results of the universality class of dipolar ferromagnets \cite{aharony1973critical,fisher1973dipolar}, since the static behaviour of the $\vpsi$ field in SMG belongs to this class.
From now on, if not explicitly indicated, we shall work with the Fourier transformed fields, defined by the relation
\begin{equation}
\psi_\alpha\left(\vx\right)=\int_{\vk}\eu^{\iu \vk\cdot\vx} \psi_\alpha\left(\vk\right)
\label{eq:fourierK}
\end{equation}
where we introduced the notation $\int\limits_{\vk}=\int\limits_{\abs{\vk}<\Lambda}\frac{d^d k}{\left(2\pi\right)^d}$.

In Eq.~\eqref{eq:fourierK} the cutoff $\Lambda$ is the value of the wave-vector above which fluctuations have no physical meaning.
Since a field theory, in the context of biophysics, is obtained by coarse-graining a discrete model, fluctuations cannot occure on distances shorter than the microscopic length $a$, therefore meaning that the cutoff can be taken to be $\Lambda\sim a^{-1}$.
In Fourier space the solenoidal constraint reads 
\begin{equation}
k_\alpha\psi_\alpha\left(\vk\right)=0
\label{eq:solenoidal}
\end{equation}
An equivalent way in which this can be formulated is by requiring that
\begin{equation}
P_{\alpha\beta}\left(\vk\right)\psi_\beta\left(\vk\right)=\psi_\alpha\left(\vk\right)
\end{equation}
where we have introduced the transverse projection operator
\begin{equation}
P_{\alpha\beta}\left(\vk\right)=\delta_{\alpha\beta}-\frac{k_\alpha k_\beta}{k^2}
\label{eq:P}
\end{equation}

The static behaviour of a field characterized by the Hamiltonian of the SSS model, given in Eq.~\eqref{eq:HamSSS}, to which a solenoidal constraint is applied, is described by the static universality class of isotropic dipolar ferromagnets \cite{aharony1973critical,bruce1974critical}.
The universality class of dipolar ferromagnets describes the more generic class of critical theories described by an Hamiltonian with a structure as that of Eq.~\eqref{eq:HamSSS}, in which the $\psi_\parallel\left(\vk\right)$ mode is taken to be non-critical.

\begin{table}[h!]
\caption{Values of $2\nu$ and $\eta$ up to order $\varepsilon^2$ \cite{bruce1974critical}}
\label{tab:critical}
\centering
\begin{tabular}[c]{c c c c c c}
 & & \multicolumn{2}{c}{Landau-Ginzburg} & \multicolumn{2}{c}{Dipolar Ferromagnets}\\
 & Mean-field & $\varepsilon$-expansion & $\varepsilon=1$ & $\varepsilon$-expansion & $\varepsilon=1$\\
\noalign{\smallskip}\hline\noalign{\smallskip}
$2\nu$ & $1$ & $1+\frac{1}{4}\varepsilon+\frac{1}{8}\varepsilon^2$ & $1.375$ & $1+\frac{9}{34}\varepsilon+\frac{7013}{58956}\varepsilon^2$ & $1.384$\\
\noalign{\smallskip}\hline\noalign{\smallskip}
$\eta$ & $0$ & $\frac{1}{48}\varepsilon^2$ & $0. 0208$ & $\frac{20}{867}\varepsilon^2$ & $0. 0231$\\
\noalign{\smallskip}\hline
\end{tabular}
\end{table}

The renormalization group analysis of isotropic dipolar ferromagnets \cite{aharony1973critical} shows that the non-criticality of the $\psi_\parallel\left(\vk\right)$ mode leads to its full suppression in the long-wavelength limit, meaning that the RG stable fixed point describes a solenoidal-constrained theory. To order  $\varepsilon = 4-d$, the RG recursive relation for the ferromagnetic coupling constant is \cite{aharony1973critical},
\begin{equation}
u_{l+1} =b^{\varepsilon}u_l \left[1-\frac{17}{2} {u}_l\ln{b}\right]
\label{zonko}
\end{equation}
In dimension $d<4$ the stable fixed point  $u^*$ ruling the critical behaviour of dipolar ferromagnets is given by \cite{aharony1973critical}
\begin{equation}
u^* =\frac{2}{17}\varepsilon 
\end{equation}
(for the sake of simplicity we have set to 1 the volume of the $d$-dimensional unit sphere).
The fact that at the stable fixed point the $\psi_\parallel\left(\vk\right)$ mode is suppressed \cite{aharony1973critical,bruce1974critical} means that the solenoidal theory is robust with respect to weak violations of the constraint, i.e. small non-critical fluctuations of the $\psi_\parallel\left(\vk\right)$ mode.
Even though the change of static universality class is very interesting at the theoretical level, it must be said that the new critical exponents are so close to those of the Landau-Ginzburg universality class that the difference is often experimentally difficult to observe \cite{bruce1974critical}.
A two-loop estimation of the critical exponents $2\nu$, ruling the behaviour of the mass $r$ while the transition is approached, and $\eta$, modifying the spatial dependence of the correlation function, are reported in Table~\ref{tab:critical}.
Here it is clear the little difference between the scaling behaviour of dipolar ferromagnets class and Landau-Ginzburg class.

\subsection{Dynamics of the solenoidal theory}\label{sec:dynamics}

In this crucial section, the dynamical equations of the mode-coupling theory subject to a solenoidal constraints will be derived. We will use the Poisson relations, the effective Hamiltonian of the SSS model, and the classic Mori-Zwanzig formalism \cite{zwanzig1961memory,mori1974new}.

Because the field $\vpsi$ is subject to the constraint,
\begin{equation}
k_\alpha\psi_\alpha\left(\vk\right)=0
\end{equation}
the most natural representation in which one would like to derive the equations of motion is,
\begin{equation}
\vpsi\left(\vk\right)= \psi_{\parallel} \left(\vk\right)\ve^{\parallel}\left(\vk\right) + \sum_{i=1}^{d-1}\psi_{\perp,i} \left(\vk\right)\ve^{\perp,i}\left(\vk\right)
\label{eq:coord}
\end{equation}
where $\ve^{\parallel}=\vk/\abs{\vk}$ is the unitary vector identifying the direction of $\vk$, while the $\ve^{\perp,i}$ are orthogonal unitary vectors spanning the space perpendicular to $\vk$, in such a way that $\ve^{\parallel}\cdot\ve^{\perp,i}=0$ and $\ve^{\perp,i}\cdot\ve^{\perp,j}=\delta_{ij}$; indeed, within this set of coordinates, the constraint simply reads,
\begin{equation}
\psi_{\parallel}\left(\vk\right)=0
\end{equation}
leaving only the $(d-1)$ independent modes $\psi_{\perp,i}\left(\vk\right)$ to take care of.
The advantage of this notation is that we can formulate the constrained field theory in terms of the $(d-1)$ independent transverse modes and not in terms of a constrained set of $d$ cartesian coordinates, $\psi_\alpha$, to which it is not clear how to apply the Mori-Zwanzig procedure. However, the explicit form of both the effective hamiltonian $\Ham$ and of the Poisson-brackets {\it are} given in terms of the cartesian coordinates, $\psi_\alpha$, while their form in terms of the $\psi_{\perp,i}$ would be extremely cumbersome. What we shall do, then, will be to first obtain the equation of motions for the $\psi_{\perp,i}$ and then to go back to the standard $\psi_\alpha$ fields by using the chain rule. In doing that, something new will pop out in the spin equation.

The dynamic behaviour of the constrained variables $\psi_\alpha$ is given, in terms of the independent variables $\psi_\perp$, by
\begin{equation}
\pder{\psi_\alpha}{t}\left(\vk,t\right)= \sum_{i=1}^{d-1}\he_\alpha^{\perp,i}\left(\vk\right)\pder{\psi_{\perp,i}}{t} \left(\vk,t\right)
\label{eq:coord_sol}
\end{equation}
Following the Mori-Zwanzig procedure, the equations of motion for the $(d-1)$ independent fields $\psi_{\perp,i}$ and $\vs$ take the following form
\begin{widetext}
\begin{gather}
\pder{\psi_{\perp,i}}{t} \left(\vk,t\right) = - \Gamma_0 \dder{\Ham}{\psi_{\perp,i}\left(-\vk\right)} + \frac{1}{2} \int_{\vq}\left\{s_{\gamma\nu}\left(-\vq\right),\psi_{\perp,i}\left(\vk\right)\right\} \dder{\Ham}{s_{\gamma\nu}\left(-\vq\right)}+\he_{\beta}^{\perp,i}\left(\vk\right)\ttheta_\beta%
\label{eq:psi_perp}\\%
\pder{s_{\alpha\beta}}{t} \left(\vk,t\right) = - k^2 \Lambda_{\alpha\beta\gamma\nu}\left(\lambda_0,\vk\right) \dder{\Ham}{s_{\gamma\nu}\left(-\vk\right)} - \sum_{i=1}^{d-1}\int_{\vq} \left\{s_{\alpha\beta}\left(\vk\right),\psi_{\perp,i}\left(-\vq\right)\right\} \dder{\Ham}{\psi_{\perp,i}\left(-\vq\right)}+ \zeta_{\alpha\beta}%
\label{eq:s_perp}%
\end{gather}
\end{widetext}
where the $t$ dependence is always understood even when not made explicit for reasons of space.
In Eqs.~\eqref{eq:psi_perp} and \eqref{eq:s_perp} $\Ham$ is the Hamiltonian given by Eq.~\eqref{eq:HamSSS} in which $\psi_\parallel$ is set to $0$; $\vecc{\ttheta}$, $\vecc{\zeta}$ are white gaussian noises with variance respectively given by $2\Gamma_0\delta_{\alpha\beta}$ and $4 k^2 \Lambda_{\alpha\beta\gamma\nu}$.
The tensor $\Lambda_{\alpha\beta\gamma\nu}=\Lambda_{\alpha\beta\gamma\nu}\left(\lambda_0;\vk\right)$ is function of the diffusive coefficient $\lambda_0$ and potentially also of the wave-vector $\vk$; in the field theory without constraint it takes the form $\Lambda_{\alpha\beta\gamma\nu}=\lambda_0 \II_{\alpha\beta\gamma\nu}$; however, since the solenoidal constraint generates an anisotropy in Fourier space for the order parameter, we expect that anisotropic effects can affect also the spin dynamics;
therefore, we generalize $ \Lambda_{\alpha\beta\gamma\nu}$ by allowing $\lambda_0$ to take different values for the longitudinal and transverse components, namely by taking,
\begin{equation}
\Lambda_{\alpha\beta\gamma\nu}\left(\lambda_0^\perp,\lambda_0^\parallel;\vk\right)=\lambda_0^{\perp}\mathbb{P}_{\alpha\beta\gamma\nu}\left(\vk\right) + \lambda_0^{\parallel}\left[\mathbb{I}_{\alpha\beta\gamma\nu}-\mathbb{P}_{\alpha\beta\gamma\nu}\left(\vk\right)\right]
\end{equation}
where $\mathbb{P}$ is the generalization of the projection operator $P$ defined in Eq.~\eqref{eq:P} acting on the space of 2-indices antisymmetric tensors, and takes the following form
\begin{equation}
\mathbb{P}_{\alpha\beta\gamma\nu}\left(\vk\right)=\II_{\alpha\beta\gamma\nu}-\II_{\alpha\beta\sigma\tau}P_{\sigma\gamma}\left(\vk\right)P_{\tau\nu}\left(\vk\right)
\end{equation}
In order to find the equations of motion of $\psi_\alpha$ and $\vs_{\alpha\beta}$, we will proceed by making explicit the terms in Eqs.~\eqref{eq:psi_perp} and \eqref{eq:s_perp} exploiting the chain rule between the fields $\psi_{\perp}$ and $\psi_\alpha$.
The chain rule applied to the variations of the Hamiltonian with respect to $\psi_{\perp,i}$ reads,
\begin{widetext}
\begin{equation}
\dder{\Ham}{\psi_{\perp,i}\left(-\vk\right)}=
\int d^dp \; \dder{\psi_\beta\left(-\vp\right)}{\psi_{\perp,i}\left(-\vk\right)}\dder{\Ham}{\psi_\beta\left(-\vp\right)}=
\he_{\beta}^{\perp,i}\left(\vk\right)\dder{\Ham}{\psi_\beta\left(-\vk\right)}
\label{eq:H_perp2psi}
\end{equation}
while the Poisson-bracket relation between $\vs$ and $\psi_{\perp,i}$ can be written as
\begin{equation}
\left\{s_{\gamma\nu}\left(-\vq\right),\psi_{\perp,i}\left(\vk\right)\right\}=
\int d^dp\left\{s_{\gamma\nu}\left(-\vq\right),\psi_\beta\left(\vp\right)\right\}\dder{\psi_{\perp,i}\left(\vk\right)}{\psi_\beta\left(\vp\right)}=\he_\beta^{\perp,i}\left(\vk\right)\left\{s_{\gamma\nu}\left(-\vq\right),\psi_\beta\left(\vk\right)\right\}
\label{eq:poisson_perp2psi}
\end{equation}
where in the last equality of both Eq.~\eqref{eq:H_perp2psi} and \eqref{eq:poisson_perp2psi} we used the relations,
\begin{align}
\dder{\psi_{\perp,i}\left(\vk\right)}{\psi_\beta\left(\vp\right)}&=\he_\beta^{\perp,i}\left(\vk\right)  \delta^{\left(d\right)}\left(\vk-\vp\right) &
\dder{\psi_\beta\left(\vp\right)}{\psi_{\perp,i}\left(\vk\right)}&=\he_\beta^{\perp,i}\left(\vk\right)  \delta^{\left(d\right)}\left(\vk-\vp\right) 
\end{align}
Thanks to Eq.~\eqref{eq:H_perp2psi} and \eqref{eq:poisson_perp2psi} we can write the mode-coupling term of the dynamics of the spin in the following way
\begin{equation}
\begin{split}
\sum_{i=1}^{d-1}\int_{\vq}\left\{s_{\alpha\beta}\left(\vk\right),\psi_{\perp,i}\left(-\vq\right)\right\}\dder{\Ham}{\psi_{\perp,i}\left(-\vq\right)}
&=\int_{\vq} \left\{s_{\alpha\beta}\left(\vk \right),\psi_\nu\left(-\vq\right)\right\} \left[\sum_{i=1}^{d-1} \he_\nu^{\perp,i}\left(\vq\right)\he_\rho^{\perp,i}\left(\vq\right)\right] \dder{\Ham}{\psi_\rho\left(-\vq\right)}=\\
&=\int_{\vq} \left\{s_{\alpha\beta}\left(\vk\right),\psi_\nu\left(-\vq\right)\right\} P_{\nu\rho}\left(\vq\right) \dder{\Ham}{\psi_\rho\left(-\vq\right)}
\end{split}
\label{eq:ham_s_perp2psi}
\end{equation}
where we used the following relation, which comes directly from the definition of the unitary vectors $\ve^{\perp,i}$
\begin{equation}
P_{\alpha\beta}\left(\vk\right)=\sum_{i=1}^{d-1} \he^{\perp,i}_\alpha\left(\vk\right)\he^{\perp,i}_\beta\left(\vk\right)
\label{eq:P2e}
\end{equation}
Thanks to the relations we found in Eq.~\eqref{eq:H_perp2psi}, \eqref{eq:poisson_perp2psi} and \eqref{eq:ham_s_perp2psi} we can write Eq.~\eqref{eq:psi_perp} and \eqref{eq:s_perp} in the following form
\begin{equation}
\pder{\psi_{\perp,i}}{t}\left(\vk,t\right)  = \he_\beta^{\perp,i}\left(\vk\right)\left[ -\Gamma_0 \dder{\Ham}{\psi_\beta\left(-\vk\right)} + \frac{1}{2}\int_{\vq} \left\{s_{\gamma\nu}\left(-\vq\right),\psi_\beta\left(\vk\right)\right\} \dder{\Ham}{s_{\gamma\nu}\left(-\vq\right)}+\ttheta_\beta\right]
\label{eq:psi_perp2psi} 
\end{equation}
\begin{equation}
\pder{s_{\alpha\beta}}{t} \left(\vk,t\right) = - k^2 \Lambda_{\alpha\beta\gamma\nu} \dder{\Ham}{s_{\gamma\nu}\left(-\vk\right)} - \int_{\vq} \left\{s_{\alpha\beta}\left(\vk\right),\psi_\nu\left(-\vq\right)\right\} P_{\nu\rho}\left(\vq\right) \dder{\Ham}{\psi_\rho\left(-\vq\right)}+ \zeta_{\alpha\beta}
\label{eq:s_perp2psi} 
\end{equation}
By substituting Eq.~\eqref{eq:psi_perp2psi} in Eq.~\eqref{eq:coord_sol} and by using the explicit Poisson-bracket relation given in Eq.~\eqref{eq:poisson}, the equations of motion for the standard Cartesian components $\psi_\alpha$ and for $s_{\alpha\beta}$ can finally be written,\begin{equation}
\pder{\psi_\alpha}{t}\left(\vk,t\right) = - \Gamma_0 P_{\alpha\beta}\left(\vk\right) \dder{\Ham}{\psi_\beta\left(-\vk\right)} +
g_0 P_{\alpha\rho}\left(\vk\right) \II_{\rho\beta\gamma\nu}\int_{\vq}\psi_\beta\left(\vk-\vq\right)\dder{\Ham}{s_{\gamma\nu}\left(-\vq\right)} + \theta_\alpha
\label{eq:psi}
\end{equation}

\begin{equation}
\pder{s_{\alpha\beta}}{t} \left(\vk,t\right) = - k^2 \Lambda_{\alpha\beta\gamma\nu}\left(\vk\right) \dder{\Ham}{s_{\gamma \nu}\left(-\vk\right)} +
2\,g_0 \II_{\alpha\beta\gamma\nu} \int_{\vq} \psi_\gamma\left(\vk-\vq\right)P_{\nu\rho}\left(\vq\right)\dder{\Ham}{\psi_\rho\left(-\vq\right)}+ \zeta_{\alpha\beta} 
\label{eq:s}
\end{equation}
Here the hamiltonian $\Ham$ can be taken to be the same of SSS model, since all the terms involving derivatives of $\Ham$ with respect to $\vpsi$ are already projected.
Moreover, the two gaussian random white noises $\vecc{\zeta}$ and $\vecc{\theta}$ have variance
\begin{gather}
\langle \theta_\alpha\left(\vk,t\right) \theta_\beta \left(\vk',t'\right) \rangle = 2 \left(2\pi\right)^{d} \Gamma_0\, P_{\alpha\beta}\left(\vk\right)\,\delta^{(d)}\left(\vk+\vk'\right)\delta\left(t-t'\right)\label{eq:noise_psi}\\
\langle \zeta_{\alpha\beta}\left(\vk,t\right) \zeta_{\gamma\nu}\left(\vk',t'\right) \rangle = 4 \left(2\pi\right)^{d} \Lambda_{\alpha\beta\gamma\nu} \left(\lambda_0^\perp,\lambda_0^\parallel;\vk\right) k^2 \,\delta^{(d)}\left(\vk+\vk'\right)\delta\left(t-t'\right)\label{eq:noise_s}
\end{gather}
\end{widetext}

\subsubsection{Effects of the constraint}
The solenoidal constraint has a double effect on the equations of motion.
The first, and maybe the most trivial, is that the equation of motion for $\vpsi$ is projected orthogonally to $\vk$, as it happens in incompressible field theories with no mode-coupling interaction \cite{FNS1977,chen2015critical}.
The second effect is less obvious, and it is represented by the presence of a projection operator $P_{\nu\rho}\left(\vq\right)$ in the mode coupling interaction of the spin dynamics. The existence of this projector is a consequence of the fact that, in the presence of a solenoidal constraint, the conservative Hamiltonian force is not simply,
\begin{equation}
\mathbb{F}_\nu\left(\vq\right)= -\dder{\Ham}{\psi_\nu\left(-\vq\right)}
\end{equation}
but rather,
\begin{equation}
\mathbb{F}_\nu\left(\vq\right)= -P_{\nu\rho}\left(\vq\right)\dder{\Ham}{\psi_\rho\left(-\vq\right)}
\label{eq:force}
\end{equation}
The linear part of the force is not affected by this new projector, but the non-linear terms are.
This can be seen by writing explicitly the new force,
\begin{widetext}
\begin{equation}
\mathbb{F}_\nu\left(\vq\right)=-\left(r_0+q^2\right)\psi_\nu\left(\vq\right)+u_0 P_{\nu\rho}\left(\vq\right) \int_{\vp,\vh} \psi_\rho\left(\vp\right)\psi_\sigma\left(\vh\right)\psi_\sigma\left(\vq-\vp-\vh\right)
\end{equation}
where in the first linear term we used the fact that $P_{\nu\rho}\left(\vq\right)\psi_{\rho}\left(\vq\right)=\psi_{\nu}\left(\vq\right)$.
The linear part of the force contributes to the dynamics of $\vs$ with the same term as Model G \cite{HH1977}
\begin{equation}
\partial_t s_{\alpha\beta}\left(\vk\right)\sim g_0 \II_{\alpha\beta\gamma\nu}\int_{\vq} \left[q^2-\left(k-q\right)^2\right] \psi_\gamma\left(\vk-\vq\right) \psi_\nu\left(\vq\right)
\end{equation}
The factor $q^2-\left(k-q\right)^2$ arises as a consequence of mode-coupling and it vanishes as $\vk\to0$, thus conserving the total spin $\vecc{S}(t)=\int d^d x \,\vs\left(\vx,t\right)=\vs\left(\vk=0,t\right)$.
But now, thanks to the presence of the projector, also the non-linear term contributes to the dynamics of the spin! More precisely, it does so through a novel dynamical interaction term given by,
\begin{equation}
\partial_t s_{\alpha\beta}\left(\vk\right)\sim 2 g_0 u_0 \II_{\alpha\beta\gamma\nu}\int_{\vq,\vh,\vp} \psi_\gamma\left(\vk-\vq\right) P_{\nu\rho}\left(\vq\right) \psi_\rho\left(\vp\right)\psi_\sigma\left(\vh\right)\psi_\sigma\left(\vq-\vp-\vh\right)
\label{eq:katz}
\end{equation}
\end{widetext}
This is a completely new term, which mixes the static ferromagnetic interaction (the coupling constant $u_0$) with the dynamic mode-coupling interaction (the coupling constant $g_0$); such vertex is absent in the non-constrained theory, since when $P_{\nu\rho}$ is substituted by $\delta_{\nu\rho}$, as in the non-solenoidal case, the r.h.s. of Eq.~\eqref{eq:katz} vanishes. In honour of the historic NYC diner at Houston and Ludlow, we call this new interaction, the {\it Katz vertex}.  As we will demonstrate later on, the Katz vertex is crucial in order to keep closed and self-consistent the RG calculation and to recover the correct static critical exponents.

The Katz interaction does not vanish when $\vk\to0$, meaning that the equation of motion of $\vs$ cannot be written as a continuity equation anymore and thus that the total spin $\vecc{S}(t)$ is no longer conserved. However, in the following section we will show thats no spin dissipation is generated by the Katz vertex, suggesting that the violation of the spin conservation is equivalent to a generalized precession of the total spin vector.

\subsection{Field theoretical description}

In order to set up the renormalization of the theory we will follow the procedure proposed by Martin, Siggia, Rose \cite{martin1973statistical}, Janssen \cite{janssen1976on} and De Dominicis \cite{de1976techniques} to write stochastic differential equations as a field theory formulated using path integrals.
Thanks to this procedure, the behaviour of a field $\vecc{\phi}$ governed by a stochastic differential equation with a deterministic evolution operator $\boldsymbol{\mathcal{F}}$ and a gaussian noise $\boldsymbol{\theta}$
\begin{equation}
\boldsymbol{\mathcal{F}}\left[\boldsymbol{\phi}\right]-\boldsymbol{\theta}=0
\label{eq:stochastic_equation}
\end{equation}
can be described through a field-theoretical action that correctly reproduces the statistics, i.e. the correlation and response functions, of Eq.~\eqref{eq:stochastic_equation},
\begin{equation}
    \mathcal{S}[\boldsymbol{\hat{\phi}},\boldsymbol{\phi}]=\int_{\tvk} \left[\hat{\phi}_\alpha \mathcal{F}_\alpha\left[\boldsymbol{\phi}\right] - \hat{\phi}_\alpha L_{\alpha\beta} \hat{\phi}_\beta\right]
    \label{eq:MSRaction}
\end{equation}
where $2 L_{\alpha\beta}$ is the variance of the gaussian noise, while $\vecc{\hat{\phi}}$ is an auxiliary field.
From now on, if not explicitly mentioned, we will work in Fourier space for both the spacial and temporal dependence, s
\begin{equation}
\phi\left(\vx,t\right)=\int_{\tvk} \eu^{\iu\left(\vk\cdot\vx-\omega t\right)}\phi(\tvk)
s\end{equation}
where $\tvk=\left(\vk,\omega\right)$ and $\int_{\tvk}=\int_{\vk}\int_{-\infty}^{\infty} \frac{d\omega}{2\pi}$.
The presence of the additional field $\boldsymbol{\hat{\phi}}$ is the cost which has to be paid in order to exploit standard path integral formulation, which will allow us to use the standard rules of static renormalization and write the perturbative series in terms of Feynman diagrams.
This auxiliary field takes the name of \textit{response field}, since the propagator, namely the response function, can be written as \cite{de2006random}:
\begin{equation}
G(\tvk)=\langle\hat{\phi}(-\tvk)\phi (\tvk)\rangle
\label{eq:response}
\end{equation}
The Martin-Siggia-Rose (MSR) action for the stochastic differential equations \eqref{eq:psi} and \eqref{eq:s} takes the following form
\begin{equation}
    \mathcal{S}[\hvpsi,\vpsi,\hvs,\vs]=
    \mathcal{S}_{0,\psi}[\hvpsi,\vpsi]+
    \mathcal{S}_{0,s}[\hvs,\vs]+
    \mathcal{S}_I[\hvpsi,\vpsi,\hvs,\vs]
    \label{eq:msr}
\end{equation}
where $\mathcal{S}_{0,\psi}$ and $\mathcal{S}_{0,m}$ are the two Gaussian free action of the two fields, which reproduce the linear dynamic theory
\begin{widetext}
\begin{equation}
    \mathcal{S}_{0,\psi}
    =\int_{\tvk} \left[
    \hpsi_\alpha\left(-\tvk\right)
    \left(-\iu\omega+\Gamma_0 k^2+m_0\right)
    \psi_\alpha\left(\tvk\right)
    -\hpsi_\alpha\left(-\tvk\right)
    \Gamma_0 P_{\alpha\beta}\left(\vk\right)
    \hpsi_\beta\left(\tvk\right)
    \right]
\label{eq:S0psi}
\end{equation}
\begin{equation}
    \mathcal{S}_{0,s}
    =
    \frac{1}{2}\int_{\tvk}
    \left[
    \hs_{\alpha\beta}\left(-\tvk\right)
    \left(-\iu\omega\mathbb{I}_{\alpha\beta\gamma\nu}+k^2\Lambda_{\alpha\beta\gamma\nu}\left(\vk\right)\right)
    s_{\gamma\nu}\left(\tvk\right)
    -\hs_{\alpha\beta}\left(-\tvk\right)
    k^2 \Lambda_{\alpha\beta\gamma\nu}\left(\vk\right)
    \hs_{\gamma\nu}\left(\tvk\right)
    \right]
\label{eq:S0s}
\end{equation}
while $\mathcal{S}_I$, which takes contributes from the non-linear dynamic terms and represents the \textit{interaction} part of the action, is given by
\begin{equation}
    \begin{split}
    \mathcal{S}_I 
    =
    &
    -g_0\, \II_{\rho\beta\gamma\nu} \int_{\tvk,\tvq}
    P_{\alpha\rho}(\vk)
    \hpsi_\alpha(-\tvk)
    \psi_\beta(\tvq)
    s_{\gamma\nu}(\tvk-\tvq)
    -g_0\, \II_{\alpha\beta\gamma\nu} \int_{\tvk,\tvq}
    \vk\cdot\vq\,
    \hs_{\alpha\beta}(-\tvk)
    \psi_\gamma(-\tvq+\tvk/2)
    \psi_\nu(\tvq+\tvk/2)+
    \\
    &
    -\frac{g_0 u_0}{12}  \int_{\tvk,\tvq,\tvp,\tvh}
    K_{\alpha\beta\gamma\nu\sigma\tau}({\vk},\vq,{\vp},{\vh},{\vk}-\vq-\vp-\vh)
    \hs_{\alpha\beta}(-\tvk)
    \psi_\gamma(\tvq)
    \psi_\nu(\tvp)
    \psi_\sigma(\tvh)
    \psi_\tau(\tvk-\tvq-\tvh-\tvp)+
    \\
    &
    +\frac{J_0}{3} \int_{\tvk,\tvq,\tvp}
    Q_{\alpha\beta\gamma\nu}(\vk)
    \hpsi_\alpha(-\tvk)
    \psi_\beta(\tvq)
    \psi_\gamma(\tvp)
    \psi_\nu(\tvk-\tvq-\tvp)
    \end{split}
    \label{eq:msrI}
\end{equation}
where we have introduced the new parameters, 
\begin{align}
m_0&=\Gamma_0 r_0 & J_0=\Gamma_0 u_0
\label{eq:equilibriumParameters}
\end{align}
which are a sort of dynamical generalization of mass and ferromagnetic coupling constant.
In Eq.~\eqref{eq:msrI} the following new tensors have been introduced
\begin{equation}
Q_{\alpha\beta\gamma\nu}\left(\vk\right)=P_{\alpha\beta}\left(\vk\right)\delta_{\gamma\nu}+
P_{\alpha\gamma}\left(\vk\right)\delta_{\beta\nu}+
P_{\alpha\nu}\left(\vk\right)\delta_{\beta\gamma}
\end{equation}
\begin{equation}
\begin{split}
K_{\alpha\beta\gamma\nu\sigma\tau}\left({\vk},{\vp}_1,{\vp}_2,{\vp}_3,{\vp}_4\right)=&
\II_{\alpha\beta\gamma\rho}Q_{\rho\nu\sigma\tau}\left(\vk-\vp_1\right)+
\II_{\alpha\beta\nu\rho}Q_{\rho\gamma\sigma\tau}\left(\vk-\vp_2\right)+\\
&+\II_{\alpha\beta\sigma\rho}Q_{\rho\gamma\nu\tau}\left(\vk-\vp_3\right)+
\II_{\alpha\beta\tau\rho}Q_{\rho\gamma\nu\sigma}\left(\vk-\vp_4\right)
\end{split}
\end{equation}

\subsubsection{Free theory}\label{sec:free}
The starting point to build the perturbative expansion of the equations of motion is the free, or Gaussian, dynamic theory, obtained by setting to $0$ all the dynamic non-linear couplings, namely $g_0$ and $u_0$.
From the gaussian part of the action, given by Eqs.~\eqref{eq:S0psi} and \eqref{eq:S0s}, we can easily derive the expressions for the bare propagators and correlation functions for the effective field theory, which are given by:
\begin{align}
\langle \psi_\alpha(\tvk)\hpsi_\beta(\tvq) \rangle_0&=\mathbb{G}_{\alpha\beta}^{0,\psi}(\tvk) \tdelta(\tvk+\tvq) &
\langle s_{\alpha\beta}(\tvk) \hs_{\gamma\nu}(\tvq)\rangle_0&=\mathbb{G}_{\alpha\beta\gamma\nu}^{0,s}(\tvk)\tdelta(\tvk+\tvq)\\
\langle \psi_\alpha(\tvk)\psi_\beta(\tvq)\rangle_0&=\mathbb{C}_{\alpha\beta}^{0,\psi}(\tvk)\tdelta(\tvk+\tvq) &
\langle s_{\alpha\beta}(\tvk) s_{\gamma\nu}(\tvq)\rangle_0&=\mathbb{C}_{\alpha\beta\gamma\nu}^{0,s}(\tvk)\tdelta(\tvk+\tvq)
\end{align}
where $\tdelta(\tvh)=(2\pi)^{d+1}\delta^{(d)}(\vh)\delta(\omega_h)$.
The subscripted $0$ on thermal averages indicate that they are computed within the non-interacting theory, namely by setting $u_0=g_0=0$.
The tensors $\mathbb{G}$ and $\mathbb{C}$ are given by
\begin{gather}
\mathbb{G}_{\alpha\beta}^{0,\psi}(\tvk)=G_{0,\psi}(\tvk)\delta_{\alpha\beta}\label{eq:greenpsi}\\
\mathbb{C}_{\alpha\beta}^{0,\psi}(\tvk)=C_{0,\psi}(\tvk)P_{\alpha\beta}(\vk)\label{eq:corrpsi}\\
\mathbb{G}_{\alpha\beta\gamma\nu}^{0,s}(\tvk)=G_{0,s}^\perp(\tvk) \PP_{\alpha\beta\gamma\nu}(\vk)+G_{0,s}^\parallel(\tvk) \left[\mathbb{I}_{\alpha\beta\gamma\nu}-\PP_{\alpha\beta\gamma\nu}(\vk)\right]\label{eq:greens}\\
\mathbb{C}_{\alpha\beta\gamma\nu}^{0,s}(\tvk)=C_{0,s}^\perp(\tvk) \PP_{\alpha\beta\gamma\nu}(\vk)+C_{0,s}^\parallel(\tvk) \left[\mathbb{I}_{\alpha\beta\gamma\nu}-\PP_{\alpha\beta\gamma\nu}(\vk)\right]\label{eq:corrs}
\end{gather}
In Eq.~\eqref{eq:greenpsi}, \eqref{eq:greens}, \eqref{eq:corrpsi} and \eqref{eq:corrs} we have,
\begin{align}
G_{0,\psi}\left(\vk,\omega\right)&=\frac{1}{-\iu\omega+\Gamma_0 k^2+\Gamma_0 r_0} &
C_{0,\psi}\left(\vk,\omega\right)&=\frac{2\Gamma_0}{\omega^2+\Gamma_0^2\left(r_0+k^2\right)^2}\\
G_{0,s}^\perp\left(\vk,\omega\right)&=\frac{2}{-\iu\omega+\lambda_0^\perp k^2}&
C_{0,s}^\perp\left(\vk,\omega\right)&=\frac{4\lambda_0^\perp k^2}{\omega^2+(\lambda_0^\perp k^2)^2}\\
G_{0,s}^\parallel\left(\vk,\omega\right)&=\frac{2}{-\iu\omega+\lambda_0^\parallel k^2}&
C_{0,s}^\parallel\left(\vk,\omega\right)&=\frac{4\lambda_0^\parallel k^2}{\omega^2+(\lambda_0^\parallel k^2)^2}
\end{align}
In the diagrammatic framework, bare propagators and correlation functions are represented in the following way
\begin{align}
	\langle \psi_\alpha\hpsi_\beta\rangle_0\quad=&\quad
	\feynmandiagram [small,horizontal=c to b] {
        		c -- [fermion] b,
        	};
    	&
	\langle s_{\alpha\beta}\hs_{\gamma\nu}\rangle_0\quad=&\quad
	\feynmandiagram [small, horizontal=c to b] {
        		c -- [charged boson] b,
        	};
	\\
	\langle \psi_\alpha\psi_\beta\rangle_0\quad=&\quad
	\feynmandiagram [small, horizontal=c to b] {
        		c -- b,
        	};
    	&
    	\langle s_{\alpha\beta}s_{\gamma\nu}\rangle_0\quad=&\quad
	\feynmandiagram [small, horizontal=c to b] {
        		c -- [boson] b,
        	};
\end{align}
where the arrows in the propagators aways point in the direction of the response field.
\end{widetext}

\subsubsection{Non-linear terms: the vertices}\label{sec:nonlin}
The four terms that compose $\mathcal{S}_I$ represent the non-linear interactions in the equations of motion.
Two of them involve one field $\hvpsi$, since they derive from the equation of motion of $\vpsi$, while the other two involve one field $\hvs$, since they derive from the equation of motion of $\vs$.
In the diagrammatic framework these interactions are represented by vertices, in which different lines merge together, each representing one of the fields involved in the interaction.
Here we are representing with a solid line the fields $\psi$ and $\hpsi$, with wavy lines the fields $s$ and $\hs$.
Moreover, an arrow is used to recognize which legs represent a response field.

The first vertex - namely interaction - involving $\hvpsi$, represents the mode coupling non-linearity, 
\begin{equation}
    \begin{tikzpicture}[baseline=(a.base)]
		\begin{feynman}[small]
			\vertex (a) at (0,0) {\(\hpsi_\alpha(-\tvk)\)};
			\vertex (b) [dot] at (1.5,0) {};
			\vertex (c) at (2.25,1.3) {\(\psi_\beta(\tvq)\)};
			\vertex (d) at (2.25,-1.3) {\(s_{\gamma\nu}(\tvp)\)};
			\diagram* {
				(a) -- [fermion] (b) -- (c),
				(b) -- [boson] (d),
			};
		\end{feynman}
	\end{tikzpicture}
    =g_0 P_{\alpha\rho}\left(\vk\right)\II_{\rho\beta\gamma\nu}\tdelta(\tvk-\tvq-\tvp)
    \label{fd:mcv_vertex}
\end{equation}
This interaction represents a purely dynamic interaction, since it is proportional only to the dynamic coupling $g_0$.

The second vertex involving $\hvpsi$ derives from the ferromagnetic quartic interaction of the static hamiltonian, ensuring that the field $\vpsi$ relaxes towards the static equilibrium distribution, and therefore is proportional to static coupling $u_0$.
It is represented by the term
\begin{equation}
	\begin{tikzpicture}[baseline=(a.base)]
		\begin{feynman}[small]
			\vertex (a) at (0,0) {\(\hpsi_\alpha(-\tvk)\)};
		    \vertex (b)[square dot] at (1.5,0) {};
			\vertex (c) at (2.25,1.3) {\(\psi_\beta(\tvq)\)};
			\vertex (d) at (2.75,0) {\(\psi_{\gamma}(\tvp)\)};
			\vertex (e) at (2.25,-1.3) {\(\psi_{\gamma}(\tvh)\)};
			\diagram* {
				(a) -- [fermion] (b) -- (c),
				(b) -- (d),
				(b) -- (e),
			};
		\end{feynman}
	\end{tikzpicture}
    	=-\frac{J_0}{3}Q_{\alpha\beta\gamma\nu}\left(\vk\right) \tdelta(\tvk-\tvq-\tvp-\tvh)
    \label{fd:ferro_vertex}
\end{equation}

The other two vertices involve one field $\hvs$, and both derive from the mode-coupling interaction in the equation for $\vs$.
The first comes from the linear part of the "force" defined in Eq.~\eqref{eq:force}, representing a purely dynamic interaction proportional to $g_0$, and it takes the usual form as in the non-constrained theory
\begin{equation}
	\begin{tikzpicture}[baseline=(a.base)]
		\begin{feynman}[small]
			\vertex (a) at (0,0) {\(\hs_{\alpha\beta}(-\tvk)\)};
			\vertex (b)[empty dot] at (1.7,0) {};
			\vertex (c) at (2.45,1.3) {\(\psi_\gamma(\tvq)\)};
			\vertex (d) at (2.45,-1.3) {\(\psi_\nu(\tvp)\)};
			\diagram* {
				(a) -- [charged boson] (b) -- (c),
				(b) -- (d),
			};
		\end{feynman}
	\end{tikzpicture}
    	=\frac{g_0}{2}\,\left(p^2-q^2\right)\,\II_{\alpha\beta\gamma\nu}\tdelta(\tvk-\tvq-\tvp)
    \label{fd:mcs_vertex}
\end{equation}
Here the factor $(p^2-q^2)$, coming from the cross product structure of the mode-coupling interaction, guarantees that this interaction vanishes when $\vk=0$.
This is a consequence of the symmetry of the non-constrained theory, which conserves the total instantaneous spin $\vecc{S}(t)=\vs(\vk=0,t)$.

The last interaction term is the Katz vertex, given by the novel non-linear contribution peculiar of the solenoidal theory, discussed at the end of Sec.~\ref{sec:dynamics}. This interaction mixes static and dynamic terms, since it represents the effects of the static quartic interaction on the dynamics of $\vs$, mediated by the mode-coupling dynamic interaction. Therefore, the Katz vertex is proportional to the product of the static coupling $u_0$ and the dynamic coupling $g_0$ and it takes the following form,
\begin{widetext}
\begin{equation}
		\begin{tikzpicture}[baseline=(a.base)]
		\begin{feynman}[small]
			\vertex (a) at (0,0) {\(\hs_{\alpha\beta}(-\tvk)\)};
			\vertex (b) [crossed dot] at (1.7,0) {};
			\vertex (c) at (1.3,1.425) {\(\psi_\gamma(\tvq_1)\)};
			\vertex (d) at (2.91,0.88) {\(\psi_\nu(\tvp_2)\)};
			\vertex (e) at (2.91,-0.88) {\(\psi_\sigma(\tvp_3)\)};
			\vertex (f) at (1.3,-1.425) {\(\psi_\tau(\tvp_4)\)};
			\diagram* {
				(a) -- [charged boson] (b) -- (c),
				(b) -- (d),
				(b) -- (e),
				(b) -- (f),
			};
		\end{feynman}
	\end{tikzpicture}
    		=\frac{g_0 u_0}{12} K_{\alpha\beta\gamma\nu\sigma\tau}\left({\vk},\vp_1,\vp_2,\vp_3,\vp_4\right)\tdelta(\tvk-\tvp_1-\tvp_2-\tvp_3-\tvp_4)
    	\label{fd:mixed_vertex}
\end{equation}
\end{widetext}
The Katz vertex arises as a consequence of the solenoidal constraint, since the tensor $K_{\alpha\beta\gamma\nu\sigma\tau}$ vanishes in the non-constrained theory. At variance with the vertex \eqref{fd:mcs_vertex}, Katz causes a violation of the spin conservation due to the fact that the order parameter has lost the $O(d)$ symmetry as a consequence of the solenoidal constraint, meaning that it does not vanish when $\vk=0$.
However, the fact that the total spin is not conserved does not mean that it is dissipated. In fact, we will show in the following sections that no perturbative corrections dissipating the spin arise after the shell integration as a consequence of the presence of the Katz vertex, reinforcing the hypothesis according to which $\vs$ is an hydrodynamic slow-variable of the system. Note that, were this not the case, the RG flow would lead to the stable fixed point of solenoidal Model A \cite{cavagna2019short}.


\section{Renormalization Group calculation}\label{sec:RG}

The key idea behind the renormalization group is that, under the assumption of an infinitely large correlation length, scaling laws and critical exponents can be obtained by looking at how the parameterss of a theory change by changing the length-scale at which the system is observed \cite{wilson1971renormalization1}.
The RG itself consists in a transformation through which a set of equations describing the dependence of the couplings from the length-scale, namely the $\beta$-functions, can be derived.

We will use Wilson's momentum shell approach \cite{wilson1971renormalization2}, which can be performed by following a two-steps procedure: \textit{i)} integrating out the short wavelength details, hence decreasing the cutoff in momentum space; \textit{ii)} rescaling space and time, so to formally reinstate the same original cutoff.
The first step is carried out by marginalizing the probability distribution of the fields over the modes in the momentum shell $\Lambda/b<k<\Lambda$, where $\Lambda$ is the cutoff.
Here $b$ is a parameter larger than $1$, but close to it, while $\Lambda$ is the \textit{cutoff} of the theory.
The effect of this integration is to modify the parameters of the theory, and shift the cutoff from $\Lambda$ to $\Lambda/b$.
To compare the coefficients of the new theory with the bare ones, space, and consequently time, must be rescaled in order to restore the original cutoff.
The iteration of this procedure defines a \textit{flow} in the parameter space, from which information on the critical behaviour can be obtained.

In the Gaussian theory, namely when $u_0=g_0=0$, the shell integration is harmless since modes at different wavelength are independent \cite{goldenfeld_lectures_1992}; hence, only the rescaling step remains, giving to all parameters, fields, and coupling constants their naive (or engineering) scaling dimension.
However, in the interacting theory, where non-linear interactions are present, during the shell integration the coupling between long and short wavelength modes generates perturbative corrections to the parameters of the theory, which have the effect to correct the scaling dimensions.
In order to explicitly compute these corrections, one performs a perturbative expansion of the shell integrals in terms of the parameter $\varepsilon=d_c-d$, where $d_c$ is the upper critical dimension, namely the dimension above which mean-field theory is exact.
This expansion method, known as $\varepsilon$-expansion, is nowadays a well established procedure in the context of perturbative RG techniques \cite{goldenfeld_lectures_1992,parisi_book,wilson1974renormalizarion,cardy1996scaling}.

\subsection{Renormalization group equations}
In this section we will explicitly show how the renormalization group changes the parameters of the model. As stressed out before the RG procedure unfolds in two steps. In the first step we integrate out the small length scale (large momenta) fluctuations, namely fluctuations with momenta $\Lambda/b<k<\Lambda$.
The effect of this integration is twofold: \textit{i)}  it changes the cut-off of the theory, since now only fluctuation $k<\Lambda/b$ are allowed; \textit{ii)} it changes the value of the parameters of the model, which acquire corrections due to the coupling between low and high momenta fluctuations,
\begin{widetext}
\begin{equation}
    \mathcal S_{\Lambda/b} = \int \hat \psi \left[
    -i \omega  + \Gamma_0 ( 1+\delta \Gamma \ln b ) k^2 + m_0(1 +\delta m \ln b) \right] \psi + \int \hat s \left[ -i \omega + \lambda^\perp _0( 1 + \delta \lambda^\perp \ln b)k^2 + \lambda^\parallel_0 ( 1+\delta \lambda^\parallel \ln b) k^2 \right] s +\dots
\end{equation}
\end{widetext}
We omitted the tensorial structure of the action for easier reading. We remark the fact that all the corrections are proportional to the volume of the momentum shell, which is always proportional to $\ln b$. The standard way to carry out this task, and to compute the corrections to the bare parameters of the model, is using perturbation theory; to be more precise we will compute the corrections $\delta \mathcal P$ using a Feynman diagram expansion.

The second step consist in re-scaling momenta, frequencies and fields\footnote{Since we know, from the static properties of the system, that the anomalous dimension is zero at one loop order, the scaling dimensions of the fields are the naive ones \cite{HH1977}.},
\begin{align}
k&\to bk & \omega&\to b^z \omega\\ 
\psi& \to b^{d_\psi} \psi  & \hat \psi &\to b^{d_{ \hat \psi}} \hat \psi  \\ 
s& \to b^{d_s} s  & \hat s &\to b^{d_{ \hat s}} \hat s 
\label{eq:resckw}
\end{align}
The effect of this second step is to restore the original \textit{cut-off} of the theory. Moreover after this step all the parameters of the model acquire a naive scaling factor, corresponding to their naive/engineering scaling dimension:
\begin{align}
\Gamma&\to b^{z-2}\Gamma & \lambda^\perp&\to b^{z-2}\lambda^\perp & \lambda^\parallel&\to b^{z-2}\lambda^\parallel\\
g&\to b^{z-\frac{d}{2}} g & J&\to b^{2z-d} J & m&\to b^{z} m
\end{align}
where $z$ is the dynamic critical exponents, which determines how the order parameter relaxes close to the critical point. After these two steps we completed the RG step, and we ended up with a theory defined by a \textit{new set of parameters}:
\begin{equation}
\begin{split}
    \Gamma_b & = b^{z-2} \; \Gamma_0(1+ \delta \Gamma \ln b)\\
    \lambda^\parallel & = b^{z-2} \; \lambda^\parallel _0  (1+ \delta \lambda^\parallel \ln b)\\
    \lambda^\perp_b  & =b^{z-2} \; \lambda^\perp _0  (1+ \delta \lambda^\perp \ln b)\\
    m_b &=  b^z \; m_0 ( 1+ \delta m \ln b)  \\
    g_b &= b^{z-\frac d 2} \; g_0  ( 1+ \delta g \ln b) \\ 
    J_b &= b^{2z-d} \; J_0  ( 1+ \delta J \ln b) \\ 
\end{split}
\label{eq:parameters_flow}
\end{equation}
How the parameters change iterating this procedure defines the renormalization group flow, and the fixed point of this flow rules the critical dynamics of the system. 

Even though we apparently have six equations for six parameters, it is possible to reduce the complexity of the problem by introducing an appropriate set of four effective coupling constants and effective parameters, namely, 
\begin{equation} 
f_0 = \frac{g_0^2}{\Gamma_0 \lambda_0^\parallel}  \qquad u_0= \frac{J_0}{\Gamma_0}  \qquad  w_0 = \frac{\Gamma_0}{\lambda^\parallel_0} \qquad x_0 = \frac{\lambda^\perp_0}{\lambda_0 ^\parallel}
\label{eq:effectiveParameters}
\end{equation}
that, through equations \eqref{eq:parameters_flow}, are regulated in a closed manner by the following four equations,\footnote{In fact, $u_0$ is not a new effective coupling constant! We are just going back to the original static ferromagnetic coupling constant; we are sorry for this back-and-forth, but it was quite inevitable.}
\begin{equation}
\begin{split}
    f_b &= b^\epsilon \,  f_0  \left[1+(2 \delta g -\delta \Gamma -\delta \lambda^\parallel) \ln b \right] \\ 
    u_b &= b^\epsilon \, u_0  \left[1 +(\delta J-\delta \Gamma)\ln b \right] \\ 
    w_b &= w_0 \left[1 + (\delta \Gamma-\delta \lambda^\parallel)  \ln b\right]\\
    x_b &= x_0\left[ 1+(\delta \lambda^\perp-\delta \lambda_\parallel)\ln b \right] 
\end{split}
\label{eq:flow_raw}
\end{equation}
We point out that if we set  $\lambda_0^\parallel = \lambda_0^\perp \equiv \lambda$ the set of effective parameters \eqref{eq:effectiveParameters} coincides with the one of the standard, unconstrained Model G \cite{HH1977}. Once we iterate the RG transformation, we obtain the final recursive RG flow equations, 
\begin{equation}
\begin{split}
    f_{l+1} &= b^\epsilon \,  f_{l} \left[1+(2 \delta g -\delta \Gamma -\delta \lambda^\parallel) \ln b \right] \\ 
    u_{l+1} &= b^\epsilon \, u_{l}  \left[1 +(\delta J -\delta \Gamma)\ln b \right] \\ 
    w_{l+1} &= w_{l} \left[1 + (\delta \Gamma-\delta \lambda^\parallel)  \ln b\right]\\
    x_{l+1} &= x_{l} \left[ 1+(\delta \lambda^\perp-\delta \lambda_\parallel)\ln b \right] 
\end{split}
\label{zanzibar}
\end{equation}
The fixed point value of these parameters will determine the value of the critical exponents of the model. We notice that the naive scaling dimension of the effective couplings $f_0$ and $u_0$ is $\epsilon = 4-d$, which suggest that their fixed point will be of order $\epsilon$.
In the next sections we shall compute all the contributions arising from the shell integral at first order in $\epsilon$ (one loop), using the Feynman diagrams technique.

Of course, the equations above are rather useless without a determination of the various perturbative corrections $\delta \Gamma, \delta g, \dots$, which we now calculate.

\subsection{Self-energies}

We start computing the perturbative corrections to the Gaussian parameters of the action, namely $m_0, \Gamma_0$, $\lambda^\perp_0, \lambda^\parallel _0$. Diagrammatically these perturbative corrections are given by the two-field vertex function, also referred as self-energies:
\begin{align}
	\Sigma_{\alpha\beta}(\tvk)=&
	\begin{tikzpicture}[baseline=(a.base)]
		\begin{feynman}[small]
			\vertex at (0,0) (a) {\(\hpsi_\alpha(-\tvk)\)};
			\vertex at (1.6,0) [blob] (b){};
			\vertex at (3,0) (c) {\(\psi_\beta(\tvk)\)};
			\diagram* {
				(a) -- [fermion] (b) -- (c),
			};
		\end{feynman}
	\end{tikzpicture}
	\\
	\Pi_{\alpha\beta\gamma\nu}(\tvk)=&
	\begin{tikzpicture}[baseline=(a.base)]
		\begin{feynman}[small]
			\vertex at (0,0) (a) {\(\hs_{\alpha\beta}(-\tvk)\)};
			\vertex at (1.6,0) [blob] (b){};
			\vertex at (3,0) (c) {\(s_{\gamma\nu}(\tvk)\)};
			\diagram* {
				(a) -- [charged boson] (b) --  [boson] (c),
			};
		\end{feynman}
	\end{tikzpicture}
\end{align}

The self-energies $\Sigma$ and  $\Pi$ are given by the following Feynman diagrams expansion:
\begin{widetext}
\begin{equation} 
	\Sigma_{\alpha\beta}=
    \begin{tikzpicture}[baseline=(a.base)]
		\begin{feynman}[small]
		    \vertex (a) at (0,0);
			\vertex (v1) [dot] at (1,0) {};
			\vertex (v2) [dot] at (2,0) {};
		    \vertex (b) at (3,0);
			\diagram* {
				(a) -- [fermion] (v1) -- [half left, fermion] (v2) -- (b),
				(v2) -- [half left, boson] (v1),
			};
		\end{feynman}
	\end{tikzpicture}
    +
	 \begin{tikzpicture}[baseline=(a.base)]
		\begin{feynman}[small]
		    \vertex (a) at (0,0);
			\vertex (v1) [dot] at (1,0) {};
			\vertex (v2) [empty dot] at (2,0) {};
		    \vertex (b) at (3,0);
			\diagram* {
				(a) -- [fermion] (v1) -- [half left] (v2) -- (b),
				(v1) -- [half right,charged boson] (v2),
			};
		\end{feynman}
	\end{tikzpicture}
	+
	\begin{tikzpicture}[baseline=(a.base)]
		\begin{feynman}[small]
		    \vertex (a) at (0,0);
			\vertex (v1) [square dot] at (1,0) {};
			\vertex (v2) at (1,0.7);
		    \vertex (b) at (2,0);
			\diagram* {
				(a) -- [fermion] (v1) -- [out=115,in=180] (v2) -- [out=0,in=65] (v1) -- (b),
			};
		\end{feynman}
	\end{tikzpicture}
\end{equation}
\begin{equation}
	\Pi_{\alpha\beta\gamma\nu}=
    \begin{tikzpicture}[baseline=(a.base)]
		\begin{feynman}[small]
		    \vertex (a) at (0,0);
			\vertex (v1) [empty dot] at (1,0) {};
			\vertex (v2) [dot] at (2,0) {};
		    \vertex (b) at (3,0);
			\diagram* {
				(a) -- [charged boson] (v1) -- [half left, fermion] (v2) -- [boson] (b),
				(v2) -- [half left] (v1),
			};
		\end{feynman}
	\end{tikzpicture}
\end{equation}
which corrects the action as follows:
\begin{equation}
    \delta \mathcal S = \int \hat v_\alpha(-k,-\omega) \Sigma_{\alpha \beta} v_\beta(k,\omega) + \hat s_{\alpha \beta}(-k,-\omega) \Pi_{\alpha \beta \gamma \nu}(k,\omega) s_{\gamma \nu}(k,\omega)
\end{equation}
Therefore $\Sigma_{\alpha \beta}$ corrects $\Gamma_0$ and $r_0$, while $\Pi_{\alpha \beta \gamma \nu}$ corrects $\lambda_0^\perp$ and $\lambda_0 ^\parallel$.  These diagrams can be written as integrals, using the standard Feynman diagrammatic rules, and read
\begin{equation}
	\begin{split}
		\Sigma_{\alpha\beta}(\tvk)=
		&
		+g_0^2 P_{\alpha\rho}(\vk) \int_{\frac{\Lambda}{b}}^\Lambda \frac{d^dp}{(2\pi)^d} \int_{-\infty}^{\infty} \frac{d\omega}{2\pi} P_{\tau\mu}(\vp_+) \mathbb{G}_{\sigma\tau}^{0,\psi}(\tvp_+) \mathbb{C}_{\rho\sigma\mu\beta}^{0,s}(\tvp_-)\\
		&
		+g_0^2 P_{\alpha\rho}(\vk) \int_{\frac{\Lambda}{b}}^\Lambda \frac{d^dp}{(2\pi)^d} \int_{-\infty}^{\infty} \frac{d\omega}{2\pi} (k^2-p_-^2) \mathbb{G}_{\rho\sigma\tau\beta}^{0,s}(\tvp_+) \mathbb{C}_{\sigma\tau}^{0,\psi}(\tvp_-)-\\
		&
		-J_0 Q_{\alpha\beta\sigma\tau}(\vk)\int_{\frac{\Lambda}{b}}^\Lambda \frac{d^dp}{(2\pi)^d} \int_{-\infty}^{\infty} \frac{d\omega}{2\pi} \mathbb{C}_{\sigma\tau}^{0,\psi}(\tvp)
	\end{split}
\end{equation}
\begin{equation}
	\Pi_{\alpha\beta\gamma\nu}(\tvk)=
	g_0^2 \II_{\alpha\beta\sigma\mu} \II_{\rho\tau\gamma\nu} \int_{\frac{\Lambda}{b}}^\Lambda \frac{d^dp}{(2\pi)^d} \int_{-\infty}^{\infty} \frac{d\omega}{2\pi} (p_+^2-p_-^2) P_{\iota\rho}(\vp_+)\mathbb{G}_{\mu\iota}^{0,\psi}(\tvp_+) \mathbb{C}_{\sigma\tau}^{0,\psi}(\tvp_-)
\end{equation}
\end{widetext}
where $\tvp_+=\tvp+\frac{\tvk}{2}$, while $\tvp_-=\tvp-\frac{\tvk}{2}$. 
The integration over the frequency is performed explicitly, as $\omega$ has no cutoff, while the integration in the wave vector is performed by using a thin-shell approximation, which is valid for $b\sim 1$; in this way we get:
\begin{equation}
\begin{split}
\Sigma_{\alpha \beta} (\tvk) =& -(m_0+\Gamma_0 k^2) \delta_{\alpha \beta} \frac{3 (1+3 w_0 +2 x_0) f_0 }{4(1+w_0)(x_0+w_0)}\ln{b} +\\
&+\delta_{\alpha \beta} \frac 9 2 u_0 (m_0-\Gamma_0\Lambda^2 ) \ln{b}
\end{split}
\label{eq:self_energyS}
\end{equation}
\begin{equation}
\begin{split}
\Pi_{\alpha \beta \gamma \nu}(\tvk)=& - \lambda_0 ^\perp k^2 \PP_{\alpha \beta \gamma \nu}(\vk)  \frac{f_0}{6x_0}  \ln{b} -\\
&-\lambda_0 ^\parallel k^2  \left[\mathbb{I}_{\alpha \beta \gamma \nu}-\PP_{\alpha \beta \gamma \nu}(\vk)\right] \frac {f_0}{3}  \ln{b}
\end{split}
\label{eq:self_energyP}
\end{equation}
Here we evaluate $\Sigma_{\alpha \beta}$ and $\Pi_{\alpha \beta \gamma \nu}$ a the relevant order in the momenta, namely up to order $k^2$. From \eqref{eq:self_energyS}, \eqref{eq:self_energyP} it is possible to read the perturbative correction to the parameters $\Gamma_0,m_0, \lambda_0 ^\perp , \lambda_0 ^\parallel$: 
\begin{align}
        \delta \Gamma &= \frac{3 (1+3 w_0+ 2 x_0)}{4(1+w_0)(x_0+w_0)} f_0 
        \label{amleto}
        \\ 
        \delta m \ &= -\frac 9 2 u_0 \frac{m_0-\Gamma_0\Lambda^2}{m_0}+\frac{3 (1+3 w_0+ 2 x_0)}{4(1+w_0)(x_0+w_0)} f_0  \\ 
        \delta \lambda^\perp &= \frac{f_0}{6 x_0}\label{eq:dll}\\ 
        \delta \lambda^\parallel & = \frac{f_0}{3}\label{eq:dlp}
\end{align}

\subsubsection{Absence of spin dissipation}

It is important at this point to emphasize a key result: the self-energy $\Pi_{\alpha \beta \gamma \nu}(k,\omega)$ vanishes when $k\to 0$. This fact implies that the renormalization group is not generating a dissipative term for the spin, namely it is not generating a linear term in the spin equation of motion which is finite at $k=0$. This result is strictly related to the particular structure of the mode coupling vertex of the spin equation of motion \eqref{fd:mcs_vertex}, and in particular to the fact that this vertex vanishes at zero external momentum $k$. We believe that this result, which we proved here only at one loop level, could be valid at all orders in perturbation theory. Indeed, the most general diagram that can generate a dissipation is given by,
\begin{equation}
\Pi_{\alpha\beta\gamma\nu}=
\begin{tikzpicture}[baseline=(a.base)]
\begin{feynman}[small]
   \vertex (a) at (0,0);
\vertex (v1) [blob] at (1,0) {};
\vertex (v2) [dot] at (2,0) {};
   \vertex (b) at (3,0);
\diagram* {
(a) -- [charged boson] (v1) -- [half left, fermion] (v2) -- [boson] (b),
(v2) -- [half left] (v1),
};
\end{feynman}
\end{tikzpicture}
\end{equation}
where the blob represents the sum of all 1-particle irriducible diagrams compatible with the given external legs, namely the renormalized mode coupling spin vertex, in which all the possible diagrammatic corrections (at all orders) are taken into account. If the renormalized spin mode coupling vertex vanishes at zero external momentum $k$, then this diagram is zero too at $k = 0$, implying that no dissipation is generated. Therefore, as far as the structure of the spin mode coupling vertex is preserved under RG, no spin dissipation can be generated. Even though in this work we explicitly showed that the structure of this vertex is preserved under RG only at one loop level, there are some hints suggesting that this result could remain valid at all order in perturbation theory.

It can be shown, by computing the Fokker-Planck equation \cite{HH1977} from the Langevin equations \eqref{eq:psi} and \eqref{eq:s}, that the probability density of the system approaches the equilibrium Gibbs-Boltzmann stationary state, $P\sim\exp\left(-\Ham\right)$. This non-perturbative result relies on the specific structure of the mode coupling vertices, which means that if the structure of the mode coupling interactions \eqref{fd:mcs_vertex} and \eqref{fd:mcv_vertex} were different, the system would not have a stationary equilibrium distribution. As this result is not perturbative, it is reasonable to think that the RG does not violate it; indeed, it would be very strange if under coarse-graining an equilibrium model flowed to an out-of-equilibrium one. Therefore, we believe that the structure of the mode coupling vertices is preserved by the RG at all orders in perturbation theory. Finally, since an essential requirement to generate a spin dissipation is that the RG changes the structure of the mode coupling vertices, we believe that no dissipation is generated at all orders in perturbation theory. Note also that this result ensures that in the presence of a small bare dissipation, the same crossover observed in \cite{cavagna2019short,cavagna2019long} between a conservative and a dissipative dynamics is recovered.

Finally, we remark once again that the absence of spin dissipation does {\it not} imply that the total spin is instantaneously conserved: due to the presence of the novel Katz vertex \eqref{eq:katz}, which is not zero at $k=0$, conservation is broken, even though it is still true that the {\it mean} value of the total spin vector is conserved, suggesting that a generalized spin precession occurs: similarly to a standard angular momentum  without dissipation but in presence of an external force, which performs a periodic precession with constant time average, so the total spin in presence of the solenoidal constraint has a non-dissipative time dynamics that conserves its mean value.

\begin{widetext}
\subsection{Mode coupling vertex corrections}

The corrections to the mode coupling constant $g_0$ are given by the following vertex functions,
\begin{align}
	V_{\alpha\beta\gamma\nu}^{\hpsi\psi s}(\tvk,\tvq)&=
	\begin{tikzpicture}[baseline=(a.base)]
		\begin{feynman}[small]
			\vertex (a) at (0,0) {\(\hpsi_\alpha(-\tvk)\)};
			\vertex (b) [blob] at (1.5,0) {};
			\vertex (c) at (2.25,1.3) {\(\psi_\beta(\tvq)\)};
			\vertex (d) at (2.25,-1.3) {\(s_{\gamma\nu}(\tvk-\tvq)\)};
			\diagram* {
				(a) -- [fermion] (b) -- (c),
				(b) -- [boson] (d),
			};
		\end{feynman}
	\end{tikzpicture}
&
    	V_{\alpha\beta\gamma\nu}^{\hs\psi\psi}(\tvk,\tvq)&=
	\begin{tikzpicture}[baseline=(a.base)]
		\begin{feynman}[small]
			\vertex (a) at (0,0) {\(\hs_{\alpha\beta}(-\tvk)\)};
			\vertex (b)[blob] at (1.7,0) {};
			\vertex (c) at (2.45,1.3) {\(\psi_\gamma(\frac{\tvk} 2 -\tvq)\)};
			\vertex (d) at (2.45,-1.3) {\(\psi_\nu(\frac{\tvk} 2 +\tvq)\)};
			\diagram* {
				(a) -- [charged boson] (b) -- (c),
				(b) -- (d),
			};
		\end{feynman}
	\end{tikzpicture}
\end{align}
These two vertex functions correct the action as follows:
\begin{equation}
    \delta \mathcal S =  \int_{\tvk,\tvq} \hat \psi_\alpha(-\tvk)\psi_\beta(\tvq) s_{\gamma\nu}(\tvk-\tvq) V^{\hat \psi \psi s}_{\alpha \beta \gamma \nu}(\tvk,\tvq)+\int_{\tvk,\tvq} \hat s_{\alpha \beta}(-\tvk) \psi _\gamma  ( \tvk/2- \tvq  ) \psi_\nu \ ( \tvk/2+ \tvq ) V^{\hat s \psi \psi}_{\alpha \beta \gamma \nu} (\tvk,\tvq)
\end{equation}
\end{widetext}
We are interested in how  these vertex function change the value of the mode coupling constant, hence we can  compute $V^{\hat \psi \psi s}$ at the zeroth order in the momenta, and $V^{\hat s \psi \psi}$ at the second order in the momenta (because the $g_0 \hat s \psi \psi$ term in the action is of second order in the wave vector).

\subsubsection*{A first key consistency check: perturbative renormalization vs symmetry generator}

The coupling $g_0$ is the parameter conjugated to the generator of the rotational symmetry (the spin), hence it plays a central role in the definition of the Poisson structure and for this reason it cannot take perturbative contributions from the RG calculation. Therefore, we expect both $V^{\hat \psi \psi s}$ and $V^{\hat s \psi \psi}$ to be zero, and fortunately this is indeed the case in our calculation. From the technical point of view, however, the fact that these vertex functions are zero is extremely nontrivial and it is worth showing, as it is a vital consistency check of the calculation and in particular of the necessity of the new Katz vertex. The vertex function $V^{\hat \psi \psi s}$ is given by the following Feynman diagrams:
\begin{widetext}
\begin{equation}
V^{\hat \psi \psi s}_{\alpha \beta \gamma \nu} =
\begin{tikzpicture}[baseline=(a.base)]
	\begin{feynman}[small]
		\vertex (a) at (0,0) ;
   		\vertex (b) at (1.9,1.125);
    		\vertex (c) at(1.9,-1.125);
		\vertex (v1)[dot] at (0.6,0){};
    		\vertex (v2)[black, dot] at (1.6,0.625){};
    		\vertex (v3) [black, dot] at (1.6,-0.625){};
    		\diagram*{
			(a) --[fermion](v1)--[ fermion](v2)--[fermion](v3)--[boson](v1);
    			(v2)--[boson](b);
			(v3)--[black](c)
    		};
	\end{feynman}
\end{tikzpicture}
+
\begin{tikzpicture}[baseline=(a.base)]
	\begin{feynman}[small]
		\vertex (a) at (0,0) ;
   		\vertex (b) at (1.9,1.125);
    		\vertex (c) at(1.9,-1.125);
		\vertex (v1)[dot] at (0.6,0){};
    		\vertex (v2)[black, dot] at (1.6,0.625){};
    		\vertex (v3) [black, empty dot] at (1.6,-0.625){};
    		\diagram*{
			(a) --[fermion](v1)--[fermion](v2)--[black](v3)--[anti charged boson](v1);
    			(v2)--[boson](b);
			(v3)--[black](c)
    		};
	\end{feynman}
\end{tikzpicture}
+
\begin{tikzpicture}[baseline=(a.base)]
	\begin{feynman}[small]
		\vertex (a) at (0,0) ;
   		\vertex (b) at (1.9,1.125);
    		\vertex (c) at(1.9,-1.125);
		\vertex (v1)[dot] at (0.6,0){};
    		\vertex (v2)[black, dot] at (1.6,0.625){};
    		\vertex (v3) [black, empty dot] at (1.6,-0.625){};
    		\diagram*{
			(a) --[fermion](v1)--[black](v2)--[anti fermion](v3)--[anti charged boson](v1);
    			(v2)--[boson](b);
			(v3)--[black](c)
    		};
	\end{feynman}
\end{tikzpicture}
+\ 
\begin{tikzpicture}[baseline=(v1.base)]
	\begin{feynman}[small]
		\vertex (a) at (0,0.7);
		\vertex (b) at (0,-0.7);
		\vertex (v1) [square dot] at (0.4,0) {};
		\vertex (v2) [dot] at (1.4,0) {};
		\vertex (c) at (2,0);
		\diagram* {
			(a) -- [fermion] (v1) -- [half left,fermion] (v2) -- [boson] (c),
			(b) -- (v1) -- [half right] (v2),
		};
	\end{feynman}
\end{tikzpicture}
\end{equation}
\end{widetext}
To compute the corrections to $g_0$ we must compute $V^{\hat \psi \psi s}$ at the zeroth order it the momenta. As it happens in the non-constrained case \cite{HH1977}, these four diagrams, at zero external momenta $\tvk$, $\tvq$, sum up to zero.
\begin{equation}
    V^{\hat \psi \psi s}_{\alpha \beta \gamma \nu}(0,0)= 0
\end{equation}
The vertex function $V^{\hat s \psi \psi}_{\alpha \beta \gamma \nu}(\tvk,\tvq)$, on the other hand, is given by the following nonzero Feynman diagrams:
\begin{widetext}
\begin{equation}
V^{\hat s \psi \psi}_{\alpha \beta \gamma \nu } = 
\begin{tikzpicture}[baseline=(a.base)]
	\begin{feynman}[small]
		\vertex (a) at (0,0) ;
   		\vertex (b) at (1.9,1.125);
    		\vertex (c) at(1.9,-1.125);
		\vertex (v1)[empty dot] at (0.6,0){};
    		\vertex (v2)[black, dot] at (1.6,0.625){};
    		\vertex (v3) [black, empty dot] at (1.6,-0.625){};
    		\diagram*{
			(a) --[charged boson](v1)--[fermion](v2)--[charged boson](v3)--[black](v1);
    			(v2)--[black](b);
			(v3)--[black](c)
    		};
	\end{feynman}
\end{tikzpicture}
+\ \begin{tikzpicture}[baseline=(a.base)]
	\begin{feynman}[small]
		\vertex (a) at (0,0) ;
   		\vertex (b) at (1.9,1.125);
    		\vertex (c) at(1.9,-1.125);
		\vertex (v1)[empty dot] at (0.6,0){};
    		\vertex (v2)[black, dot] at (1.6,0.625){};
    		\vertex (v3) [black, empty dot] at (1.6,-0.625){};
    		\diagram*{
			(a) --[charged boson](v1)--[fermion](v2)--[boson](v3)--[anti fermion](v1);
    			(v2)--[black](b);
			(v3)--[black](c)
    		};
	\end{feynman}
\end{tikzpicture}
+\ 
\begin{tikzpicture}[baseline=(v1.base)]
	\begin{feynman}[small]
		\vertex (a) at (0,0);
		\vertex (v1) [empty dot] at (0.6,0) {};
		\vertex (v2) [square dot] at (1.6,0) {};
		\vertex (b) at (2,0.8);
		\vertex (c) at (2,-0.8);
		\diagram* {
			(a) -- [charged boson] (v1) -- [half left,fermion] (v2) -- (b),
			(c) -- (v2) -- [half left] (v1),
		};
	\end{feynman}
\end{tikzpicture}
+ \ 
\begin{tikzpicture}[baseline=(v1.base)]
	\begin{feynman}[small]
		\vertex (a) at (0,0);
		\vertex (v1) [crossed dot] at (1,0) {};
		\vertex (v2) at (1,1);
		\vertex (b) at (1.7,0.8);
		\vertex (c) at (1.7,-0.8);
		\diagram* {
			(a) -- [charged boson] (v1) -- [out=115,in=180] (v2) -- [out=0,in=65] (v1) -- (b),
		        (v1) -- (c),
		};
	\end{feynman}
\end{tikzpicture}
\end{equation}
\end{widetext}
The first two diagrams (the triangles) cancel each other, exactly as in the non-solenoidal case. The other two diagrams are, on the other hand, specific to the solenoidal case.  The first diagram, which vanishes in the non-constrained theory \cite{cavagna2019long}, is nonzero when the solenoidal constraint is present, due the suppression of the longitudinal $\psi$ mode,
\begin{equation}
    \begin{tikzpicture}[baseline=(v1.base)]
\begin{feynman}[small]
\vertex (a) at (0,0);
\vertex (v1) [dot] at (1,0) {};
\vertex (v2) [square dot] at (2,0) {};
\vertex (b) at (2.5,0.8);
\vertex (c) at (2.5,-0.8);
\diagram* {
	(a) -- [charged boson] (v1) -- [half left,fermion] (v2) -- (b),
	(c) -- (v2) -- [half left] (v1),
};
\end{feynman}
\end{tikzpicture}
	=-\frac{g_0 u_0}{8} \II_{\alpha\beta\sigma\tau}\left(k_\sigma k_\gamma \delta_{\tau\nu} +k_\sigma k_\nu \delta_{\tau\gamma}\right)+\dots
	\label{fd:bubble}
\end{equation}
where the ellipses stand for higher order in the momentum expansion, representing corrections to RG-irrelevant interactions.  
This vertex corrections not only would give to $g_0$ a perturbative correction due to the shell integration, but it would \textit{generate} a novel interaction term too, since it has not the same tensorial structure as the original interaction \eqref{fd:mcs_vertex}.
Therefore, if no other diagram canceling it were present, the RG would not have a closed structure, meaning that the equations of motion would not be eigenstates of the RG transformation, since the shell integration generates new interaction terms that were not present in the bare theory. This unpleasant scenario, in which new relevant interactions arise during the RG flow, is avoided by the key presence of a new Feynman diagram formed by a bubble connection of two lines of the Katz vertex, namely, 
\begin{equation}
    \begin{tikzpicture}[baseline=(v1.base)]
    \begin{feynman}[small]
    \vertex (a) at (0,0);
    \vertex (v1) [crossed dot] at (1,0) {};
    \vertex (v2) at (1,1);
    \vertex (b) at (1.7,0.8);
    \vertex (c) at (1.7,-0.8);
    \diagram* {
    	(a) -- [charged boson] (v1) -- [out=115,in=180] (v2) -- [out=0,in=65] (v1) -- (b),
    	(v1) -- (c),
    };
    \end{feynman}
    \end{tikzpicture}
	=\frac{g_0 u_0}{8} \II_{\alpha\beta\sigma\tau}\left(k_\sigma k_\gamma \delta_{\tau\nu} +k_\sigma k_\nu \delta_{\tau\gamma}\right)+\dots
\end{equation}
where, as before, the ellipses stand for higher order in the momentum expansion.
The presence of this diagram is fundamental, since it {\it exactly cancels} the contributions of diagram \eqref{fd:bubble}, therefore curing the anomalies that the latter carries and making the solenoidal RG calculation self-consistent. We therefore consider the following diagrammatic equation a key result of our calculation:
\begin{equation}
\boxed{
	\begin{split}
	\\
   \qquad\begin{tikzpicture}[baseline=(v1.base)]
\begin{feynman}[small]
\vertex (a) at (0,0);
\vertex (v1) [dot] at (1,0) {};
\vertex (v2) [square dot] at (2,0) {};
\vertex (b) at (2.5,0.8);
\vertex (c) at (2.5,-0.8);
\diagram* {
	(a) -- [charged boson] (v1) -- [half left,fermion] (v2) -- (b),
	(c) -- (v2) -- [half left] (v1),
};
\end{feynman}
\end{tikzpicture} +
\begin{tikzpicture}[baseline=(v1.base)]
\begin{feynman}[small]
\vertex (a) at (0,0);
\vertex (v1) [crossed dot] at (1,0) {};
\vertex (v2) at (1,1);
\vertex (b) at (1.7,0.8);
\vertex (c) at (1.7,-0.8);
\diagram* {
	(a) -- [charged boson] (v1) -- [out=115,in=180] (v2) -- [out=0,in=65] (v1) -- (b),
	(v1) -- (c),
};
\end{feynman}\qquad 
\end{tikzpicture} =0 \qquad  \\ 
\\
\end{split}
}
\end{equation}
Moreover, although the Katz vertex contributes to the dynamic behaviour of $\vs$ also at $\vk=0$, namely violating the conservation of the total spin, the last diagrams vanishes at $\vk=0$. The fact that the simplest one-loop diagram that can be constructed starting from the Katz vertex does not give any perturbative contribution at vanishing momenta suggests that no spin dissipation should arise, not even at higher orders in the $\varepsilon$-expansion.
This is related to the fact that the Katz vertex violates the conservation of the spin leading to a precession of it, but keeping fixed its average value.

\subsection{Ferromagnetic vertex corrections} 
In contrast to the mode coupling vertex, the ferromagnetic coupling does have perturbative corrections due to the shell integration.  The coupling $J_0$ is corrected by the four-field vertex function $V^{\hat \psi\psi \psi \psi}$,
\begin{widetext}
\begin{equation}
    	V_{\alpha\beta\gamma\nu}^{(\hpsi\psi \psi \psi)}(\tvk,\tvq,\tvp)=
	\begin{tikzpicture}[baseline=(a.base)]
		\begin{feynman}[small]
			\vertex (a) at (0,0) {\(\hpsi_\alpha(-\tvk)\)};
		    \vertex (b)[blob] at (1.5,0) {};
			\vertex (c) at (2.25,1.3) {\(\psi_\beta(\tvq)\)};
			\vertex (d) at (2.75,0) {\(\psi_{\gamma}(\tvp)\)};
			\vertex (e) at (2.25,-1.3) {\(\psi_{\nu}(\tvk-\tvq-\tvp)\)};
			\diagram* {
				(a) -- [fermion] (b) -- (c),
				(b) -- (d),
				(b) -- (e),
			};
		\end{feynman}
	\end{tikzpicture} 
\end{equation}
that corrects the action as follows:
\begin{equation}
    \delta \mathcal S = \int_{\tvk,\tvq,\tvp} \hat \psi_\alpha(-\tvk) \psi_\beta(\tvp) \psi_\gamma(\tvq) \psi_\nu(\tvk-\tvq-\tvp) \ V^{\hat \psi \psi \psi \psi}_{\alpha \beta \gamma \nu}(\tvk,\tvq,\tvp)
\end{equation}
At one loop, only the following Feynman diagrams contribute to the vertex function $V^{\hat \psi \psi \psi \psi}$ in a non-trivial way:
\begin{equation}
        V_{\alpha\beta\gamma\nu}^{\hpsi\psi \psi \psi}=
        \begin{tikzpicture}[baseline=(v1.base)]
		\begin{feynman}[small]
		    \vertex (a) at (0,0.5);
		    \vertex (b) at (0,-0.5);
			\vertex (v1) [square dot] at (0.5,0) {};
			\vertex (v2) [square dot] at (1.5,0) {};
			\vertex (c) at (2,0.5);
			\vertex (d) at (2,-0.5);
			\diagram* {
				(a) -- [fermion] (v1) -- [half left] (v2) -- (c),
				(d) -- (v2) -- [half left] (v1) -- (b),
			};
		\end{feynman}
	\end{tikzpicture}+
	\begin{tikzpicture}[baseline=(v1.base)]
		\begin{feynman}[small]
		    \vertex (a) at (0,0);
			\vertex (v1) [dot] at (1,0) {};
			\vertex (v2) [crossed dot] at (2,0) {};
		    \vertex (b) at (2.5,0.7);
			\vertex (c) at (2.7,0);
			\vertex (d) at (2.5,-0.7);
			\diagram* {
				(a) -- [fermion] (v1) -- [half left,charged boson] (v2) -- (b),
				(v2) -- [half left] (v1),
				(v2) -- (c),
				(v2) -- (d),
			};
		\end{feynman}
	\end{tikzpicture}
	+\begin{tikzpicture}[baseline=(v2.base)]
		\begin{feynman}[small]
		    \vertex (a) at (0,1);
		    \vertex (b) at (0,-1);
		    \vertex (v1) [dot] at (0.3,0.5) {};
			\vertex (v2) [square dot] at (1.1,0) {};
			\vertex (v3) [dot] at (0.3,-0.5) {};
			\vertex (c) at (1.6,0.5);
			\vertex (d) at (1.6,-0.5);
			\diagram* {
				(a) -- [fermion] (v1) -- [fermion] (v2) -- [fermion] (v3) -- [boson] (v1) ,
				(b) -- (v3),
				(c) -- (v2) -- (d),
			};
		\end{feynman}
	\end{tikzpicture}+
	\begin{tikzpicture}[baseline=(v2.base)]
		\begin{feynman}[small]
		    \vertex (a) at (0,1);
		    \vertex (b) at (0,-1);
		    \vertex (v1) [dot] at (0.3,0.5) {};
			\vertex (v2) [square dot] at (1.1,0) {};
			\vertex (v3) [empty dot] at (0.3,-0.5) {};
			\vertex (c) at (1.6,0.5);
			\vertex (d) at (1.6,-0.5);
			\diagram* {
				(a) -- [fermion] (v1) -- [fermion] (v2) -- (v3) -- [anti charged boson] (v1) ,
				(b) -- (v3),
				(c) -- (v2) -- (d),
			};
		\end{feynman}
	\end{tikzpicture}+
	\begin{tikzpicture}[baseline=(v2.base)]
		\begin{feynman}[small]
		    \vertex (a) at (0,1);
		    \vertex (b) at (0,-1);
		    \vertex (v1) [dot] at (0.3,0.5) {};
			\vertex (v2) [square dot] at (1.1,0) {};
			\vertex (v3) [empty dot] at (0.3,-0.5) {};
			\vertex (c) at (1.6,0.5);
			\vertex (d) at (1.6,-0.5);
			\diagram* {
				(a) -- [fermion] (v1) --  (v2) -- [anti fermion] (v3) -- [anti charged boson] (v1) ,
				(b) -- (v3),
				(c) -- (v2) -- (d),
			};
		\end{feynman}
	\end{tikzpicture}
    \label{eq:diagramsFerro}
\end{equation}
The first term is the classic fish diagram of the standard ferromagnetic theory; the second diagram is generated by joining a mode-coupling vertex with the Katz vertex; the last three diagrams are of purely mode-coupling origin. Computing $V ^{\hat \psi\psi\psi\psi}$  at the zeroth order in the external momenta, gives,
\begin{equation}
    V^{\hat \psi \psi \psi \psi}  _{\alpha \beta \gamma \nu} =  -\frac{J_0}{3}  Q_{\alpha \beta \gamma \nu} \left[- \frac{17}{2} u_0 +\frac{3 (1+3 w_0 +2 x_0)}{4(1+w_0)(x_0+w_0)} f_0  \right] \ln b
\end{equation}
\end{widetext}
so that the perturbative corrections to the coupling $J_0$ are,
\begin{equation}
\delta J  =- \frac{17}{2} u_0 +\frac{3 (1+3 w_0 +2 x_0)}{4(1+w_0)(x_0+w_0)} f_0 
\label{bonzo}
\end{equation}

\subsubsection{A second key consistency check: statics vs dynamics} 
The dynamical RG calculation (at equilibrium) must of course contain in itself the static RG calculation; more specifically, if a coupling constant is present also in the static case, its dynamical renormalization must be exactly the same as its static renormalization. This is clearly the case for the ferromagnetic coupling, which is perfectly well-defined also within a purely static framework. Therefore, in this section we show that this consistency between statics and dynamics is achieved by our calculation.

First of all we recall that the {\it actual} ferromagnetic coupling, namely the coupling constant that appears in the static Hamiltonian, is $u_0= J_0 / \Gamma_0$ (see \eqref{eq:equilibriumParameters}). Hence, the static ferromagnetic coupling $u_0$ gets perturbative corrections both from $J_0$ and $\Gamma_0$,
\begin{equation}
    \delta u = \delta J -\delta \Gamma 
\end{equation}
and from equations \eqref{bonzo} and \eqref{amleto} we have, 
\begin{equation}
    \delta u = -\frac{17}{2} u_0
    \label{suino}
\end{equation} 
Hence, the static coupling $u_0$ does {\it not} receive any perturbative corrections from the dynamic coupling $g_0$, which is healthy.
But the crucial check is whether the recursive relation we get for $u$ from the dynamic RG is the same as the static one, equation \eqref{zonko}. Fortunately, it is.
From equation \eqref{zanzibar} and \eqref{suino} we obtain,
\begin{equation}
u_{l+1} = b^\varepsilon \, u_{l}  \left[1 -\frac{17}{2} u_{l} \ln b \right] 
\end{equation}
which is exactly the same as the static RG recursive equation \eqref{zonko}. We stress that this key consistency is recovered in an extremely nontrivial way; in particular, the Katz vertex plays a crucial role. The cancellation of the dynamical coupling $g_0$ in the perturbative correction of the static coupling is achieved through the following diagrammatic identity:
\begin{widetext}
\begin{equation}
\boxed{
\begin{split}
&\\ \quad 
	&\begin{tikzpicture}[baseline=(v1.base)]
		\begin{feynman}[small]
		    \vertex (name) at (1,1) {$D_1$};
		    \vertex (a) at (0.5,0);
			\vertex (v1) [dot] at (1,0) {};
			\vertex (v2) [crossed dot] at (2,0) {};
		    \vertex (b) at (2.2,0.7);
			\vertex (c) at (2.3,0);
			\vertex (d) at (2.2,-0.7);
			\diagram* {
				(a) -- [fermion] (v1) -- [half left,charged boson] (v2) -- (b),
				(v2) -- [half left] (v1),
				(v2) -- (c),
				(v2) -- (d),
			};
		\end{feynman}
	\end{tikzpicture} \ +
	\begin{tikzpicture}[baseline=(v2.base)]
		\begin{feynman}[small]
		    \vertex (name) at (1,1) {$D_2$};
		    \vertex (a) at (0,1);
		    \vertex (b) at (0,-1);
		    \vertex (v1) [dot] at (0.3,0.5) {};
			\vertex (v2) [square dot] at (1.1,0) {};
			\vertex (v3) [dot] at (0.3,-0.5) {};
			\vertex (c) at (1.4,0.5);
			\vertex (d) at (1.4,-0.5);
			\diagram* {
				(a) -- [fermion] (v1) -- [fermion] (v2) -- [fermion] (v3) -- [boson] (v1) ,
				(b) -- (v3),
				(c) -- (v2) -- (d),
			};
		\end{feynman}
	\end{tikzpicture}+
	\begin{tikzpicture}[baseline=(v2.base)]
		\begin{feynman}[small]
		\vertex (name) at (1,1) {$D_3$};
		    \vertex (a) at (0,1);
		    \vertex (b) at (0,-1);
		    \vertex (v1) [dot] at (0.3,0.5) {};
			\vertex (v2) [square dot] at (1.1,0) {};
			\vertex (v3) [empty dot] at (0.3,-0.5) {};
			\vertex (c) at (1.4,0.5);
			\vertex (d) at (1.4,-0.5);
			\diagram* {
				(a) -- [fermion] (v1) -- [fermion] (v2) -- (v3) -- [anti charged boson] (v1) ,
				(b) -- (v3),
				(c) -- (v2) -- (d),
			};
		\end{feynman}
	\end{tikzpicture}+
	\begin{tikzpicture}[baseline=(v2.base)]
		\begin{feynman}[small]
		    \vertex (name) at (1,1) {$D_4$};
		    \vertex (a) at (0,1);
		    \vertex (b) at (0,-1);
		    \vertex (v1) [dot] at (0.3,0.5) {};
			\vertex (v2) [square dot] at (1.1,0) {};
			\vertex (v3) [empty dot] at (0.3,-0.5) {};
			\vertex (c) at (1.4,0.5);
			\vertex (d) at (1.4,-0.5);
			\diagram* {
				(a) -- [fermion] (v1) --  (v2) -- [anti fermion] (v3) -- [anti charged boson] (v1) ,
				(b) -- (v3),
				(c) -- (v2) -- (d),
			};
		\end{feynman}
	\end{tikzpicture} =  \frac{\partial}{\partial k^2}
		\begin{tikzpicture}[baseline=(a.base)]
		\begin{feynman}[small]
		    \vertex (name) at (0.8,1) {$D_5$};
			\vertex at (0,0) (a) ;
			\vertex at (0.7,0) [blob] (b){};
			\vertex at (1.3,0) (c) ;
			\diagram* {
				(a) -- [fermion] (b) --  [black] (c),
			};
		\end{feynman}
	\end{tikzpicture} \\ 
	&\\
	&\qquad \qquad D_1 = - \frac{f_0}{4(w_0+x_0)}\\ 
	&\qquad \qquad D_2 +D_3 + D_4 = \frac{(2+5 w_0+3 x_0)}{2(1+w_0)(x_0+w_0)}f_0 \\ 
	&\qquad  \qquad D_5  = \frac{3 (1+3 w_0 +2 x_0)}{4(1+w_0)(x_0+w_0)} f_0 \\
	&
		\end{split} 
		}
    \label{eq:diagramsFerrobis}
\end{equation}
\end{widetext}
where the l.h.s of the equation is computed at zero external momenta. The diagram $D_1$ is the product of the interplay between the mode-coupling vertex and the Katz vertex, which is therefore crucial in recovering the correct static behaviour.

There is a second, and subtler, consistency check related to the renormalization of the ferromagnetic coupling constant. The coupling $u_0$ not only appears in front of the ferromagnetic vertex, but - due to the static-dynamic coupling induced by the solenoidal constraint - it also appears in front of the Katz vertex, $V^{\hat s  \psi \psi \psi \psi }$, which is indeed proportional to $g_0 u_0$ (see equation \eqref{fd:mixed_vertex}); we know that $g_0$ does not acquire perturbative corrections, hence any diagrammatic correction to the Katz vertex must be billed to $u_0$; but $u_0$ has been already corrected by its natural ferromagnetic vertex $V^{\hat \psi \psi \psi \psi}$, in the static-compliant way that we have just seen, equation \eqref{suino}. Hence, it seems we have two potentially {\it independent} corrections to $u_0$, one coming from the {\it bona fide} ferromagnetic vertex, and a second one from the Katz vertex! If these diagrammatic corrections were different from each other, we would have a serious problem, as there would be a bifurcation of the ferromagnetic interaction, with highly dubious physical interpretation, not to mention the impossible recovery of the equilibrium static results. Once again, fortunately, the calculation does not disappoint us, even though in a very nontrivial way. The Katz vertex function $V^{\hat s \psi\psi\psi\psi}$ has (at one loop) the following non-vanishing diagrammatic contribution,
\begin{equation}
    V_{\alpha\beta\gamma\nu\sigma\tau}^{\hs\psi\psi\psi\psi}=
    \begin{tikzpicture}[baseline=(v1.base)]
		\begin{feynman}[small]
		    \vertex (a) at (0,0);
		    \vertex (c) at (0.4,0.6);
		    \vertex (b) at (0.4,-0.6);
			\vertex (v1) [crossed dot] at (0.7,0) {};
			\vertex (v2) [square dot] at (1.7,0) {};
			\vertex (d) at (2.2,0.5);
			\vertex (e) at (2.2,-0.5);
			\diagram* {
				(a) -- [charged boson] (v1) -- [half left] (v2) -- (d),
				(e) -- (v2) -- [half left] (v1) -- (b),
				(c) -- (v1),
			};
		\end{feynman}
	\end{tikzpicture}
	\label{eq:diagramsMixed}
\end{equation}
Computing this Feynman diagram we get,
\begin{equation}
    V_{\alpha\beta\gamma\nu\sigma\tau}^{\hat s \psi\psi\psi\psi} =  -\frac{g_0 u_0}{12} \left[ 1 -\frac{17}{2} u_0 \right ]  K_{\alpha \beta \gamma \nu \sigma}
\end{equation}
and therefore the correction to $u_0$ that we obtain from the Katz vertex is exactly the same as from the ferromagnetic vertex, namely 
$\delta u = - 17/2 u_0$, which saves the day.


\section{The critical dynamics of solenoidal Model G}\label{nostromo}

We now have all that we need to finally calculate the dynamical critical exponent $z$ at one loop in a mode-coupling theory subject to a solenoidal constraint, that is in solenoidal Model G.

\subsection{The recursive RG equations and the $\beta$-functions} \label{sec:rec}

First we collect all the perturbative contributions, we plug them into equations \eqref{zanzibar} and write the recursive relations for the effective parameters and coupling constants of the theory,
\begin{align}
u_{l+1}&=b^{\varepsilon} u_l \left[1-\frac{17}{2} u_l\ln{b}\right]\\
f_{l+1}&=b^{\varepsilon} f_l \left[1-f_l\left(\frac{1}{3}+\frac{3 \left(1+3w_l+2x_l\right)}{4 \left(1+w_l\right)\left(x_l+w_l\right)}\right)\ln{b}\right]\label{eq:parF}\\
w_{l+1}&=w_l \left[1-f_l\left(\frac{1}{3}-\frac{3 \left(1+3w_l+2x_l\right)}{4 \left(1+w_l\right)\left(x_l+w_l\right)}\right)\ln{b}\right]\label{eq:parW}\\
x_{l+1}&=x_l \left[1-\frac{f_l}{3}\left(1-\frac{1}{2 x_l}\right)\ln{b}\right]\label{eq:parX}
\end{align}
where we are now working at $T=T_c$, namely at $r=m=0$.
Since the calculation here is performed at one loop, the static coupling $u$ does not contribute to the renormalization of the dynamic parameters, therefore completely decoupling the dynamic behaviour from the static one. Because we have already abundantly checked that the renormalization of $u$ is compatible with the statics, we simply drop this equation from now on. 

The derivatives of a parameter $\mathcal{P}$ with respect to $\ln{b}$ is known as the $\beta$-function $\beta_{\mathcal{P}}=\frac{\partial \mathcal{P}}{\partial\ln{b}}$, and measure how the parameter change when an infinitesimal RG transformation is performed.
The $\beta$-functions of the effective parameters $f$, $w$ and $x$, obtained from Eqs.~ \eqref{eq:parF}, \eqref{eq:parW} and \eqref{eq:parX}, are given by:
\begin{align}
\beta_f&=f\left(\varepsilon-\frac{1}{3}-\frac{3  \left(1+3w+2x\right)}{4 \left(1+w\right)\left(x+w\right)}\right)\\
\beta_w&=w f\left(\frac{3  \left(1+3w+2x\right)}{4 \left(1+w\right)\left(x+w\right)}-\frac{1}{3}\right)\\
\beta_x&=\frac{f}{3} \left(\frac{1}{2}-x\right)
\end{align}
The zeros of the $\beta$-functions give the fixed points of the RG flow, which have a crucial role in ruling the critical behaviour of the theory.
Since we are interested in a genuine mode-coupling dynamical regime, we will not consider the trivial fixed points with $f^*=0$, since they lead to the overdamped dynamics with $z= 2$ (at one loop) typical of Model A \cite{HH1977}.
We find a set of non-trivial fixed points, one stable and the other unstable.
The stable fixed point is given by,
\begin{equation}
f^* = \frac{3\varepsilon}{2} \quad , \quad w^*=\frac{21+\sqrt{697}}{8} \quad , \quad x^*=\frac{1}{2} 
\label{lucrezio}
\end{equation}
As expected the effective coupling constant is of order $\varepsilon=4-d$ at this fixed point.

\subsection{Anisotropy in the spin dynamics}
In Model G \cite{HH1977}, the absence of anisotropic interactions leads to the implicit assumption that all the different directions of the fields, both in real and Fourier space, meaning that transverse and parallel modes must be equal and therefore $\lambda=\lambda^\perp=\lambda^\parallel$, namely $x=1$.
However, at the new stable fixed point of the solenoidal Model G described by Eq.~\eqref{lucrezio} $x^*=1/2$, meaning that the anisotropy due to the suppression of the longitudinal $\psi$-mode leads to a different dynamic behaviour of the two $s$-modes $\vs^\perp$ and $\vs^\parallel$, in such a way that ${\lambda^\parallel}^*=2{\lambda^\perp}^*$.
This result directly follows from the fact that the perturbative corrections $\delta\lambda^\perp$ and $\delta\lambda^\parallel$, given in Eqs. \eqref{eq:dll} and \eqref{eq:dlp} respectively, obey the relation
\begin{equation}
\delta\lambda^\parallel=2 x \,\delta\lambda^\perp
\end{equation}
It is not yet clear to us whether this factor $2$ between $\lambda^\perp$ and $\lambda^\parallel$ can be guessed through a direct analysis of the equations of motion, or if it valid only in the long wavelength and long time dynamic behaviour.

In any case, since the diagram contributing to $\delta\lambda^\perp$ and $\delta\lambda^\parallel$ does not involve neither propagators nor correlators of the field $s$, we may observe that this result is a pure consequence of suppression of the $\psi^\parallel$ modes.
In fact, even if we had naively left $\lambda_0=\lambda_0^\perp=\lambda_0^\parallel$ at a bare level, the RG transformation would have led to two different perturbative corrections $\delta\lambda^\perp$ and $\delta\lambda^\parallel$, meaning that the infrared behaviour of this theory has two different diffusive coefficients for the $\vs^\perp$ and $\vs^\parallel$ modes.

\subsection{The dynamical critical exponent $z$}

To find the dynamic critical exponent, following \cite{HH1977} and \cite{cardy1996scaling}, we require that the kinetic coefficient of the primary field, $\Gamma$, is non-singular at the RG fixed point, thus ensuring that the effective RG theory has a non-singular characteristic time scale.
This amounts to imposing the condition,
\begin{equation}
 \Gamma^*=\lim_{l\to\infty}\Gamma_l=O(1)
 \label{nesta}
\end{equation}
and thus we need to write explicitly the recursive RG equation for the kinetic coefficient; that equation can be found in \eqref{eq:parameters_flow}, complemented by its perturbative corrections, equation \eqref{amleto}, thus giving, 
\begin{equation}
\Gamma_{l+1} = b^{z-2}\, \Gamma_l \left[
1+
\frac{
3(1+3w_l+2x_l)f_l
}{
4(1+w_l)(x_l+w_l)
}
\ln b
\right]
\label{resta}
\end{equation}
By using the obvious expansion, $b^x (1+y\ln b) = b^{x+y}$, we finally obtain from \eqref{nesta} and \eqref{resta} that the general expression for the dynamic critical exponent is,
\begin{equation}
z=2-\frac{3 \left(1+3w^*+2x^*\right) f^*}{4 \left(1+w^*\right)\left(x^*+w^*\right)}
\end{equation}
and once we use the one-loop values of the parameters at the stable fixed point \eqref{lucrezio}, we obtain
\begin{equation}
z=\frac{d}{2}
\end{equation}
Despite the difference of all effective parameters and coupling constants at the stable fixed point, this is exactly the same dynamic critical exponent as the standard unconstrained Model G \cite{HH1977}. This result is somewhat surprising. The solenoidal constraint {\it does} change the static universality class: the static critical exponents are different from the Landau-Ginzburg class, and define the novel dipolar ferromagnet class (see Table I and \cite{aharony1973critical}). According to the common wisdom in the theory of critical phenomena, we would expect a change also in the dynamic universality class, as universality is normally broader at the static level than at the dynamic level. For example, Ising-like ferromagnets, have the same static critical exponents, while the dynamical critical exponents varies depending on whether the order parameter is conserved (Model B) or not (Model A). Here, we see something different: the dynamic universality class {\it does not} change due to the solenoidal constraint. Even though we have derived this conclusion perturbatively, this is probably a non-perturbative result due to the great power of the symmetry; the lack of diagrammatic renormalization of the coupling constant conjugated to the generator of the rotations, $g_0$, leads to the recursive relation, 
\begin{equation}
g_{l+1}=b^{z-\frac{d}{2}}g_l
\end{equation}
If we now ask that the central charge of the symmetry does not change {\it at all}, we get $z=d/2$, at the non-perturbative level.

The fact that the solenoidal constraint does not alter the dynamical critical exponent is an encouraging result, since it suggests that using 
the incompressibility condition to simplify the dynamical equations of active matter with mode coupling interactions is a reasonable approximation, as it does not change dramatically the dynamical behaviour of the system.


\section{Conclusions}
We have studied the effects of a solenoidal constraint on the critical dynamics of a field $\vpsi$ with $O(d)$-symmetry in the presence of  mode-coupling interaction with the generator of the rotational symmetry $\vs$, which we called spin; more succinctly, we have studied solenoidal Model G.
The presence of the constraint leads to the suppression of the $\psi$-mode parallel to the wave-vector $\vk$, namely $\psi^\parallel(\vk)=\vk\cdot\vpsi\left(\vk\right)/\abs{k}$, violating the $O(d)$-symmetry and modifying the static behaviour.
The equations of motion of the constrained theory have been derived starting from the symmetries and Poisson-bracket relations between the hydrodynamic variables, namely the order parameter $\vpsi$ and the spin $\vs$.
We performed a one-loop renormalization group calculation to investigate the long wave-length and long time behaviour in the critical region.
The closed structure of the RG transformation and the consistency of the RG flow with the static behaviour of dipolar ferromagnets provide a self-consistent proof that no RG-relevant interaction has been omitted and that the equations of motion we derived are correct.

Two main dynamic effects arise as a consequence of the solenoidal constraint.
The first, and most predictable one, is the projection on the plane orthogonal to $\vk$ of the equation of motion for the order parameter $\vpsi$.
On the contrary, no similar projection of the equation for $\vs$ can be performed in order to obtain the spin dynamics; instead, the suppression of $\psi^\parallel(\vk)$ leads to a novel non-linear interaction - the Katz vertex - combining the effect of the static ferromagnetic coupling of the field $\vpsi$ and the mode-coupling dynamic interaction.
The presence of this new mixed interaction is the second, less intuitive, effect of the constraint, which adapts the spin dynamics to the presence of the constraint by making the static quartic coupling contributing to the torque-like interaction $\partial_t\vs\sim g_0\vpsi\times\delta_\psi\Ham$.
Moreover, the Katz vertex contributes to the time-derivative of the spin also at zero wave vector, therefore violating the conservation of the total spin.
 
The lack of conservation of the spin is not something strange: the order parameter is not rotational-invariant, due to the solenoidal constraint, and therefore the generator of its rotations is not a conserved quantity.
It is however crucial to understand whether the spin is dissipated or not, since the presence of a dissipation generated by the RG would make the spin stop being an hydrodynamic variable, therefore suppressing any inertial behaviour in the critical region.
The torque-like nature of the Katz vertex, which is the only dynamic term violating the spin conservation, indicates that this violation gives rise to a generalized precession of the total spin, rather than a dissipation.
At one loop, the perturbative expansion confirms this interpretation, since the self-energy of the spin $\Pi$ does not contain any perturbative corrections at $\vk=0$.
Moreover, we show that the presence of any dissipative term in the linear dynamics of the spin can arise only if the dynamic mode-coupling vertex in the equation of $\vs$ did not vanish when $\vk=0$, which seems not to be the case for this theory.

Our RG calculation passed several nontrivial consistency checks.
First of all, the fact that the equations of motion appear to be eigenstates of the RG, in the sense that the shell integration step does not generate new interaction terms, ensures that we did not miss any relevant coupling in the problem description.
Secondly, the RG recursive relations we found for the dynamic theory reproduce the behaviour of dipolar ferromagnets, ensuring that the static behaviour is correctly reproduced by the dynamics.
We must remark that both these results directly follow from the presence of the new Katz vertex, in absence of which the theory would not describe correctly the dynamics of the system; therefore meaning that the this new non-trivial interaction plays a crucial role in making the dynamic behaviour of the spin compatible with the constraint. 

From the study of the RG recursive relations, the dynamic behaviour has been shown to be characterized by a critical exponent $z=\frac{d}{2}$, which is the same as the non-constrained theory. This result was somewhat unexpected. In general, static properties are more robust compared to dynamical properties; models with different dynamical  critical behaviours are often characterized by the same static behaviour, such as Model A and Model G of \cite{HH1977}. On the contrary, in our model the solenoidal constraint changes the static universality class, leaving unchanged the critical dynamics of the system.
This suggests that the dynamic critical behaviour of homogeneous systems does not change when a solenoidal constraint, i.e. incompressibility, is enforced; this indicates that we can try and understand their dynamic critical behaviour by studying their incompressible version.
Homogeneous systems are governed by equations of motion in which incompressibility is, in general, not required, but where density fluctuations, and therefore density-velocity couplings, are negligible.
Incompressibility, though, completely suppresses density fluctuations and therefore represents a stronger requirement than homogeneity.
Moreover, requiring incompressibility, hence imposing a solenoidal constraint on the velocity field, generates long-range interactions that could change the properties of a system; while this is indeed the case for the static behaviour of our theory, the long-range interactions are not sufficient to modify its dynamic universality class.
Theoretical evidences of this fact have already been discussed in homogeneous active systems \cite{cavagna2020equilibrium}, therefore suggesting that the solenoidal constraint does not significantly affect the critical dynamic behaviour in the presence of neither activity nor mode-coupling terms. This result is very encouraging, as it allows to study the homogeneous phase of the off-equilibrium Inertial Spin Model under a incompressible hypothesis, where the absence of the density field leads to a great simplification of the calculation.

Our result is an important stepping stones towards developing an RG theory for natural swarms.
The Katz interaction vertex in the spin dynamics that we derived in this work, which emerges as an effect of the solenoidal constraint, must certainly characterize also the equations of motion of an incompressible out-of-equilibrium field theory, in which terms coupling the order parameter to its generator of rotations are present. It would have been extremely difficult to derive the Katz vertex had one tackled the problem directly in the presence of activity. Despite this step forward, though, the complexity of the present calculation suggests that significant more theoretical efforts will be needed to carry out the full-fledged out-of-equilibrium mode-coupling RG study of natural swarms.

We warmly thank M. Testa, C. Castellani, and U. Tauber for fruitful discussions.
L. Di Carlo also thanks I. Basile for interesting exchanges about the work.
This work was supported by ERC Advanced Grant RG.BIO (contract n.785932) to AC, and ERANET-CRIB Grant to AC and TSG. TSG was also supported by grants from CONICET, ANPCyT and UNLP (Argentina).

\bibliographystyle{apsrev4-1}
\bibliography{SMG-bibliography.bib}

\end{document}